\documentclass[%
reprint,%
floatfix,%
amsmath,amssymb,%
aps,%
pre,%
]{revtex4-2}

\usepackage{graphicx}
\usepackage{dcolumn}
\usepackage{bm}

\begin{document}

\title{Characterizing particle rearrangements in sheared highly polydisperse materials}

\author{Waad Paliwal}
\email{waad.paliwal@emory.edu}
\affiliation{Physics Department, Emory University, Atlanta, GA 30322, USA}

\author{Eric R. Weeks}
\affiliation{Physics Department, Emory University, Atlanta, GA 30322, USA}

\date{\today}

\begin{abstract}
We compare three particle-scale measures of rearrangement in highly polydisperse materials under driven flow: (a) nonaffine motion defined relative to the time-averaged mean flow, (b) changes in nearest-neighbor connectivity, and (c) $D^2_{\min}$, which measures nonaffine motion relative to an affine deformation fitted locally in space and time [Falk and Langer, \textit{Phys. Rev. E} \textbf{57}, 7192 (1998)]. We apply these measures to previously published two-dimensional simulations~\cite{JiangSussmanWeeks2023} and granular-flow experiments~\cite{IllingWeeks2025NonaffinePolydisperse} with polydispersities up to $\delta\approx0.50$. Changes in connectivity and $D^2_{\min}$ both require a definition of neighboring particles, making the choice of neighborhood nontrivial in highly polydisperse systems. For detecting changes in connectivity, we recommend radical Delaunay triangulation, which provides a size-aware topological definition of neighbors. For calculating $D^2_{\min}$, we recommend a size-aware pairwise cutoff distance method. We further show that changing the neighborhood definition can reverse the apparent dependence of $D^2_{\min}$ on particle size in experimental data. Thus, trends in $D^2_{\min}$ cannot be interpreted independently of the neighborhood used to calculate it. Overall, the three measures presented provide complementary information about rearrangements in highly polydisperse systems.
\end{abstract}

\maketitle

\section{Introduction}
\label{sec:intro}

Soft matter encompasses materials such as emulsions, foams, and
suspensions that are common in nature and industry. Their mechanical
and transport properties arise from the collective motion of discrete
constituents, including particles, droplets, and grains, particularly
under shear. In particulate soft materials, interactions among these
constituents transmit forces through a disordered contact network
~\cite{Desmond2013ForcesEmulsionDroplets}. At the microscopic scale,
deformation in disordered soft materials proceeds through particle
rearrangements and associated nonaffine motion
~\cite{Argon1979,MaloneyLemaitre2006,hebraud97,schall07,
BonnEtAl2017YieldStress}.

In a spatially varying flow, adjacent finite-sized particles are driven
at different velocities. Their interactions therefore require local
reorganizations rather than allowing each particle to independently
follow the imposed flow~\cite{losert00}. These rearrangements are
typically localized and collective rather than single-particle events
~\cite{Barrat,yamamoto97,schall07,vasisht18}, and they are spatially
heterogeneous, such that only some regions of the material rearrange at
a given time~\cite{cubuk15,hassani19}. Correlated rearrangements may
also organize into avalanche-like sequences
~\cite{lemaitre09,tsai21,jiang19}.

If the flow is reversed or varied cyclically, rearrangements may be
reversible, irreversible, or reversible with hysteresis
~\cite{hebraud97,petekidis02,lundberg08,FalkLanger1998}. We refer to
any rearrangement that is not simply reversible as a plastic
rearrangement~\cite{Tanguy2021ElastoPlastic}. Local rearrangements
redistribute stress and can also relax it~\cite{chen15,desmond15}; when
irreversible, they also dissipate energy and contribute to nonlinear
mechanical response and yielding
~\cite{BonnEtAl2017YieldStress,FalkLanger2011Review,
Tanguy2021ElastoPlastic}. Although our focus is on driven systems,
related behavior occurs in quiescent glassy materials, where thermally
driven rearrangements contribute to aging and generate spatially
heterogeneous dynamics commonly termed ``dynamical heterogeneity''
~\cite{Ediger2000SpatiallyHeterogeneous,Sillescu1999Heterogeneity,
Weeks2000StructuralRelaxation,Glotzer2000HeterogeneousDynamics,
BerthierBiroli2011GlassTransition}.

Particle-resolved simulations and microscopy make it possible to
characterize rearrangements in several ways. The first method tracks
changes in nearest-neighbor connectivity
~\cite{Rabani1997CageCorrelations,Conrad2006SlowClusters,
YamamotoOnuki1998,shiba12}. For example, in two dimensions, a ``T1
event'' occurs when two particles separate and two others come
together, thereby exchanging neighbors
~\cite{lundberg08,chen12,chen15,desmond15}.

\begin{figure}[t]
  \centering
  \includegraphics[width=0.8\columnwidth]{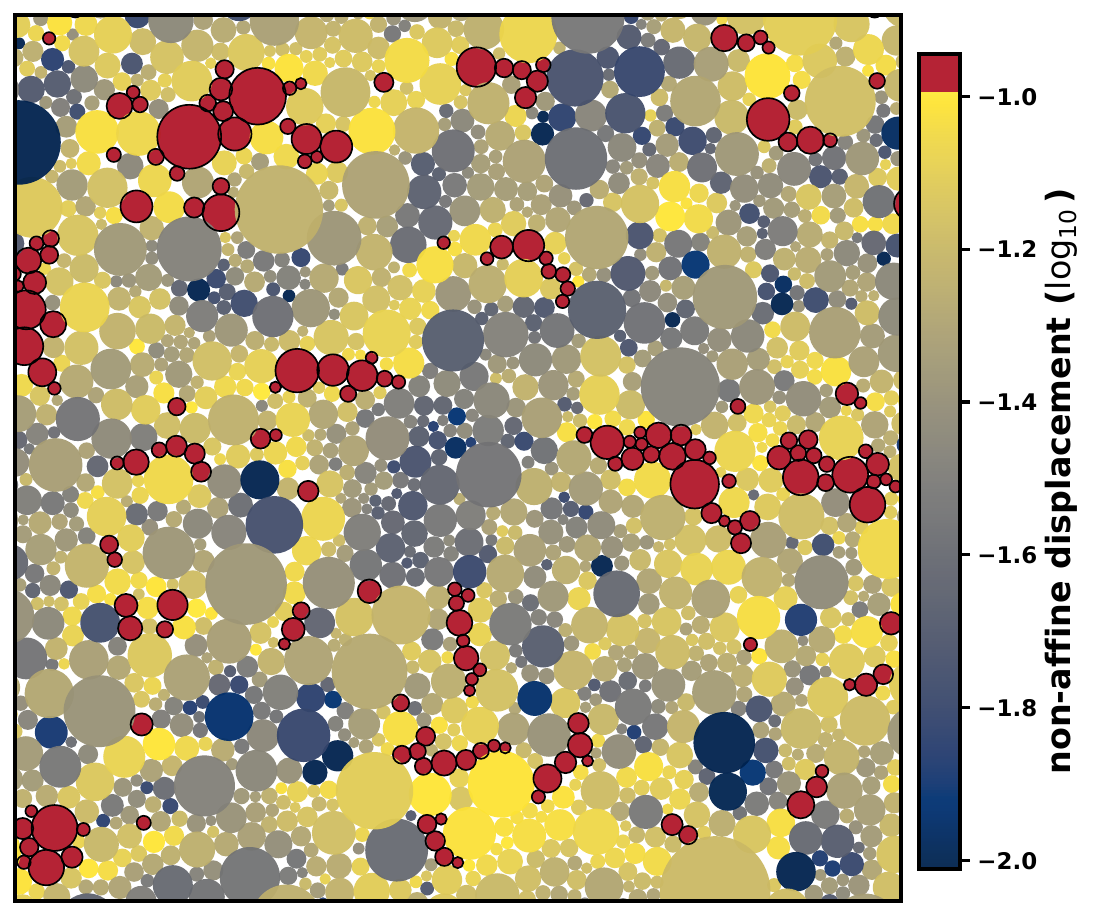}
  \caption{Nonaffine motion field in a highly polydisperse system, $\delta\sim 0.501$. Particles with large nonaffine motion are interpreted as rearranging; the deep red particles correspond to the top 10 percent. These particles appear throughout the material, illustrating the heterogeneous nature of rearrangements. They also form localized groups or clusters, highlighting the collective nature of plastic rearrangements. 
  }
  \label{fig:exp10-nonaffine-motion}
\end{figure}

The second method measures nonaffine motion relative to a mean flow.
Affine motion varies linearly with position; for example, simple shear
has velocity $\vec{v}=\dot{\gamma}y\hat{x}$, where $\dot{\gamma}$ is
the strain rate, $y$ is the transverse coordinate, and $\hat{x}$ is
the flow direction. More generally, a steady flow may have a
time-averaged velocity that varies smoothly in space. Nonaffine motion
is the component of an instantaneous particle displacement that
differs from this time-averaged flow
~\cite{liu96,YamamotoOnuki1998,Chen2010ShearedColloidalLiquid,
Utter2008AffineNonaffineGranularShear}. Such deviations arise from
interactions with neighboring particles that prevent each particle
from following the mean flow smoothly
~\cite{besseling07,Chen2010ShearedColloidalLiquid,vasisht18,
clararahola15}. Figure~\ref{fig:exp10-nonaffine-motion} illustrates the
spatial heterogeneity and collective character of this motion.

The third method, introduced by Falk and Langer
~\cite{FalkLanger1998}, measures nonaffine motion relative to an
affine transformation fitted locally in space and time. A least-squares
fit is performed using the relative displacements of a particle and
its neighbors, and the residual, $D^2_{\min}$, quantifies the motion
that cannot be described by that local affine transformation. Falk and
Langer associated regions of large $D^2_{\min}$ with
shear-transformation zones and noted that these rearrangements can be
hysteretic when the loading direction is reversed. $D^2_{\min}$ and
related particle-scale measures have since been used to identify
structural soft spots and connect local yielding thresholds to plastic
activity~\cite{manning11,cubuk15,patinet2016connecting}.

The first and third methods require a definition of ``neighbors,''
which becomes challenging when a sample contains particles spanning a
wide range of sizes. We quantify the polydispersity by $\delta$, the
standard deviation of the particle radii divided by the mean radius.
Many commonly studied model systems are monodisperse, narrowly
polydisperse, or binary mixtures with modest size contrast, including
many examples with $\delta\lesssim0.2$
~\cite{mason97emulsions,hebraud97,petekidis02,yamamoto97,teitel07,
manning11,Chen2010ShearedColloidalLiquid}. Prior work has also
considered more highly polydisperse systems
~\cite{chen15,clararahola15}, but this regime has been studied less
extensively.

Many natural and industrial materials have broad particle-size
distributions, including rock glaciers~\cite{haeberli06}, landslides and
avalanches~\cite{pitman05}, ice m{\'e}lange~\cite{burton18},
soil~\cite{or02}, mud~\cite{besq03}, cement~\cite{rosquoet03}, and
food products~\cite{taylor09}. Here we consider flowing materials with
polydispersities up to $\delta\sim0.5$ and ratios of the largest to
smallest particle radius as large as $10:1$.

Polydispersity complicates both neighbor-dependent measures. Larger
particles generally have more nearby particles, so gaining or losing
one neighbor may not have the same physical significance for large and
small particles. A single center-to-center distance cutoff is also
problematic because the appropriate separation between neighbors
depends on both particle sizes~\cite{Lynch2008AgingBinaryColloidalGlass}.
Similarly, a local affine fit requires a neighborhood that samples a
comparable physical region across particle sizes and size
distributions. The central challenge is therefore to define neighbors
without introducing artificial dependencies of the measured
rearrangements on particle size or sample composition.

In this manuscript, we first use previously published simulation data
~\cite{JiangSussmanWeeks2023} to develop and compare the three
measures. We then test the resulting interpretation using previously
published experimental data
~\cite{IllingWeeks2025NonaffinePolydisperse}. For the two
neighbor-dependent methods, we evaluate alternative definitions and
provide recommendations for highly polydisperse systems. As van Meel
\textit{et al.} emphasize, the appropriate definition of neighbors
depends on the question being asked~\cite{vanMeel2012}. Although our
recommendations are largely independent of the degree of
polydispersity, the physical question must ultimately guide both the
choice of neighborhood and the interpretation of the results.

\section{Data}

As a model system, we use the simulation results of Jiang {\it et al.}~\cite{JiangSussmanWeeks2023}. The system consists of a two-dimensional, athermal, densely packed assembly under steady shear.  The simulation implements the Durian bubble model:  the particles are soft, and in addition to their soft repulsive interactions they also have viscous interactions to mimic a foam or an emulsion~\cite{durian95,tewari99}. The particle radii $R$ are drawn from truncated exponential distributions:
\begin{equation}
P(R)\propto \exp\left(-\frac{R}{R_{\text{min}}}\right), \qquad R_{\text{min}} \le R \le R_{\text{max}}.
\end{equation}

\noindent
The size ratio $R_{\text{max}}/R_{\text{min}}$ ranges from 3 to 10. There is also a bidisperse system with size ratio $1:1.4$ and number ratio $1:1$.  Including the bidisperse sample, there are a total of five data sets with distinct polydispersities $\delta$ spanning $[0.167, 0.501]$.  For the shear protocol, the domain is a box with periodic boundary conditions in the $x$ direction and Lees--Edwards boundary conditions in $y$. The system is packed at an area fraction of $\phi = 0.93$ and subjected to a steady strain rate. Additional details of the simulation can be found in Ref.~\cite{JiangSussmanWeeks2023}.

Following Jiang's approach~\cite{JiangSussmanWeeks2023}, we use the same strain increment, $\Delta\gamma = 0.005$, to determine displacements and look for rearrangements.  We restrict our analysis to the steady-state regime. This system was chosen because it provides a clean model of a highly polydisperse material under shear, allowing us to focus on characterizing rearrangements.

We also cross-check our analysis using the experimental data reported by Illing and Weeks~\cite{IllingWeeks2025NonaffinePolydisperse}. As in the numerical simulations, the experimental system consists of highly polydisperse disks under steady flow, with polydispersity up to $\delta = 0.48$.  Additional details can be found in Ref.~\cite{IllingWeeks2025NonaffinePolydisperse}.

\section{Nonaffine motion based on the mean flow}
\label{sec:nonaffine}

We first consider nonaffine motion, defined by comparing instantaneous
particle displacements with the time-averaged flow
~\cite{YamamotoOnuki1998,Chen2010ShearedColloidalLiquid,
Utter2008AffineNonaffineGranularShear}. This is the simplest of the
three methods because it does not require a neighbor definition or
explicitly use particle sizes. In our implementation, its parameters
are the lag time $\Delta t$ and, when the mean flow must be inferred
from the data, the spatial bin size. The method also requires a
well-defined time-averaged flow, as is generally available in a
statistically steady state.

An affine displacement is the motion predicted by a smooth deformation field, such as macroscopic shear. For example, ideal parallel-plate shear of a uniform material produces a velocity profile that varies linearly across the gap. A particle displacement can therefore be decomposed into an affine component that follows the mean flow and a nonaffine component that captures instantaneous deviations from it. For a particle $i$ at position $\vec{r}_i(t)$, the displacement over a lag time $\Delta t$ is
\begin{equation}
\Delta \vec{r}_i(t)=\vec{r}_i(t+\Delta t)-\vec{r}_i(t).
\end{equation}

We determine the time-averaged displacement field, $\Delta \vec{r}_{\mathrm{mean}}(x,y)$, by averaging over time and over particles within spatial bins. If the mean flow is known independently, it need not be inferred from the particle data. For example, in the simulations analyzed here, the imposed mean flow is $\vec{v}=\dot{\gamma}y\hat{x}$. The nonaffine displacement is obtained by subtracting the mean displacement evaluated at the particle position:
\begin{equation}
\Delta \vec{r}_i^{\,\mathrm{NA}}(t)
=
\Delta \vec{r}_i(t)
-
\Delta \vec{r}_{\mathrm{mean}}\!\left(\vec{r}_i(t)\right).
\end{equation}

A nonzero value of $\Delta \vec{r}_i^{\,\mathrm{NA}}$ indicates a
departure from the mean flow. The use of nonaffine motion as a proxy
for rearrangements in amorphous materials has a long history. Under
shear, particles with large nonaffine displacements identify regions
related to the localized shear transformations described by
Argon~\cite{Argon1979}.

Figure~\ref{fig:exp10-nonaffine-motion} shows the magnitude of the
nonaffine displacement during one displacement interval in the
simulation with $\delta=0.501$. The field forms spatially correlated
regions. For particle $i$ and a lag interval beginning at time $t$, we
define
\begin{equation}
A_i(t)=\left|\Delta \mathbf{r}_i^{\mathrm{NA}}(t)\right|.
\end{equation}
Thus, $A_i(t)$ is a particle-resolved quantity evaluated separately for
each analyzed interval. For each simulation, we calculate a single
mean over all particles and all analyzed times,
\begin{equation}
\overline{A}=\left\langle A_i(t)\right\rangle_{i,t},
\end{equation}
and define the fluctuation about this global mean as
\begin{equation}
\delta A_i(t)=A_i(t)-\overline{A}.
\end{equation}
We then calculate the spatial correlation function
\begin{equation}
C_A(r)=
\frac{
\left\langle
\delta A_i(t)\,\delta A_j(t)
\right\rangle_{
\substack{
t,\,i\ne j\\
|\mathbf{r}_i(t)-\mathbf{r}_j(t)|\approx r
}}
}{
\left\langle
[\delta A_i(t)]^2
\right\rangle_{i,t}
}.
\label{eq:na_spatial_correlation}
\end{equation}
The numerator is evaluated using particle pairs at the same time whose
separations lie in the radial bin centered at $r$, and these pairs are
then pooled over all analyzed times. Numerically, this average is
estimated using randomly sampled particle pairs. We define the
correlation length $\xi$ operationally as the first distance at which
the correlation decreases through $1/e$:
\begin{equation}
C_A(\xi)=\frac{1}{e}.
\end{equation}

\begin{figure}[t]
  \centering
  \includegraphics[width=\columnwidth]{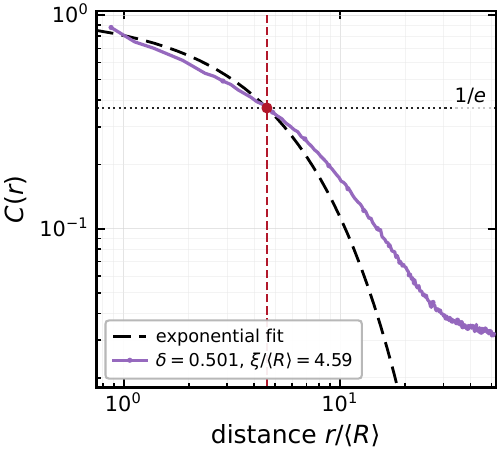}
  \caption{Log--log plot of the spatial correlation function $C_A(r)$ for the nonaffine-motion field in the sample with the highest polydispersity, $\delta=0.501$. Distance is normalized by the mean particle radius, $r/\langle R\rangle$. The horizontal dotted line marks $C_A(r)=1/e$, and the vertical dashed line indicates the corresponding correlation length, $\xi/\langle R\rangle=4.6$. An exponential fit constrained to pass through the same $1/e$ point is shown for comparison.}
  \label{fig:raw-na-global-correlation}
\end{figure}

Figure~\ref{fig:raw-na-global-correlation} shows $C_A(r)$ for a representative data set. We use the normalized connected correlation function, dividing by the
variance of $A_i$ so that $C_A(r)$ is dimensionless and $C_A(0)=1$ by construction. The violet curve bends downward on the log--log axes, indicating that the decay is not well described by a single power law. This behavior contrasts with that reported by Chikkadi and Schall, who observed power-law decay~\cite{ChikkadiSchall2012}. Their system was three-dimensional and had a lower polydispersity, $\delta=0.07$, either of which may contribute to the difference. For our data, an exponential reasonably approximates the initial decay, as shown by the dashed line. The correlation function deviates from this fit at larger $r$, where finite-size effects may become important. Applying the $1/e$ criterion across all simulation data sets gives $\xi/\langle R\rangle=4.6$--$6.6$, with no systematic dependence on $\delta$. For the bidisperse sample, $\xi/\langle R\rangle=5.9$, close to the mean value of $5.7$ across the five data sets.

An important caveat is that nonaffine motion defined in this way does not necessarily imply that particles rearrange relative to one another. A group of particles may move coherently in a manner that differs from the time-averaged flow. Particles at the boundary of that group may rearrange relative to neighbors outside the group, while particles within the group retain the same neighbors. Nonaffine motion therefore identifies regions in which particle motion departs from the mean flow, but it does not by itself establish rearrangement as defined by changes in neighbor connectivity. Nor is nonaffine motion necessarily irreversible or hysteretic. Nevertheless, particles in a homogeneous continuum undergoing steady flow at low Reynolds number would be expected to follow the mean flow; nonaffine motion is therefore a notable feature of flowing particulate materials.

\section{Neighbor swapping and bond breaking}
\label{sec:bondbreak}

A complementary topological approach to detecting rearrangements is to track changes in particle neighbors. Let $N_i(t)$ denote the set of neighbors of particle $i$ at time $t$. Local reorganization over a time or strain interval appears as gains or losses in this set. We define $\delta N_i(t;\Delta t)$ as the total number of neighbors gained or lost between $N_i(t)$ and $N_i(t+\Delta t)$, so both types of change contribute positively to $\delta N_i$. This measure follows ideas introduced by Rabani \textit{et al.}~\cite{Rabani1997CageCorrelations}. We use the terms ``neighbor swapping'' and ``bond breaking'' for these changes, although the bonds considered here are geometric connections rather than physical or chemical bonds~\cite{Rabani1997CageCorrelations,Conrad2006SlowClusters}.

Within this framework, a particle is labeled as rearranging if $\delta N_i(t;\Delta t)>0$. Applying this criterion requires a definition of neighbors. In systems with finite-range interactions, neighbors may correspond directly to chemically bonded or interacting particles. More generally, neighbor relationships can be defined geometrically. In monodisperse systems, two particles are often considered adjacent when their center-to-center separation is below a specified cutoff, commonly chosen as the first minimum of the pair correlation function $g(r)$~\cite{cianci06ssc,Lynch2008AgingBinaryColloidalGlass,ChikkadiSchall2012,Weeks2000StructuralRelaxation}. In polydisperse systems, a single absolute cutoff is less suitable because the characteristic separation between neighbors depends on the sizes of both particles.

A natural size-aware definition is provided by the radical Delaunay triangulation (RDT), which identifies neighbors topologically. To introduce RDT, first consider the ordinary Voronoi tessellation and its dual, the Delaunay triangulation, for a set of points~\cite{preparata85,Okabe2000}. The Voronoi polygon associated with each point is the region of the plane closer to that point than to any other, and these polygons together tile the plane. Two particles are considered neighbors if their Voronoi polygons share an edge~\cite{Weeks2000StructuralRelaxation,ChikkadiSchall2012}. Connecting all such neighboring pairs produces the Delaunay triangulation, which tiles the plane with triangles. In three dimensions, the corresponding structures are Voronoi polyhedra and Delaunay tetrahedra.

RDT extends this construction by assigning a size-dependent weight to each particle center. Radical Voronoi polygons are defined using the power distance from a position $\mathbf{x}$ to particle $i$,
\begin{equation}
\Pi_i(\mathbf{x})=\lVert \mathbf{x}-\mathbf{p}_i \rVert^2-R_i^2,
\label{eq:powerRD}
\end{equation}

\noindent where $\mathbf{p}_i$ and $R_i$ are the center and radius of particle $i$, respectively~\cite{Aurenhammer1987}. The radius-dependent term shifts the polygon boundaries to account for particle size. Two particles are radical Delaunay neighbors if their radical Voronoi polygons share an edge; the corresponding connection between their centers defines a bond. Because this construction accounts explicitly for particle size, it is well suited to polydisperse systems. It has been used, for example, to analyze polydisperse sphere packings~\cite{GervoisEtAl2002}, while its geometric dual, the radical Voronoi tessellation, has been used to characterize local structural heterogeneity in amorphous materials~\cite{RieserEtAl2016,du24}. The RDT neighbor relation is reciprocal: two particles are mutual neighbors whenever they share a radical Delaunay edge. The ordinary, unweighted Voronoi construction is recovered when all particle weights are set to zero in Eq.~\ref{eq:powerRD}.

We therefore define a particle as rearranging when its RDT neighbor set changes over the interval $(t,t+\Delta t)$. Although bond breaking does not by itself imply irreversibility or hysteresis, the associated rearrangements often relax local stress~\cite{desmond15}. Rearrangements may also exhibit hysteresis if the flow is reversed~\cite{lundberg08}.

\begin{figure}[t]
  \centering
  \includegraphics[width=\columnwidth]{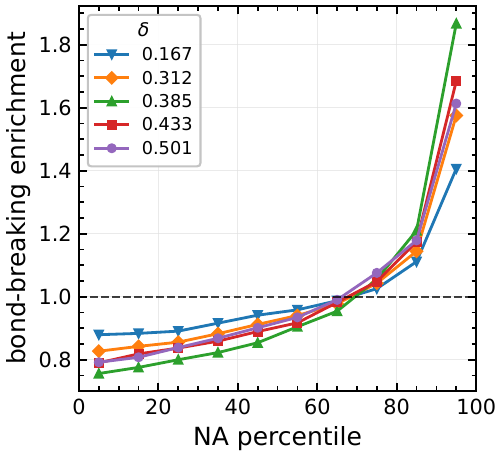}
  \caption{Bond-breaking enrichment as a function of the percentile of nonaffine-motion (NA) magnitude for the five indicated polydispersities. The dashed horizontal line at unity marks the no-enrichment baseline; values above 1 indicate that bond-breaking events are more common among particles in that range of nonaffine motion. Each point is located at the center of a decile, so the leftmost and rightmost points represent particles in the lowest and highest 10\% of nonaffine-motion magnitude, respectively.}
  \label{fig:na-percentile-bond-breaking}
\end{figure}

Not surprisingly, bond breaking and nonaffine motion are related. To quantify this relationship, we calculate an enrichment factor: the probability that a particle gains or loses a neighbor, conditioned on its nonaffine-motion decile, divided by the corresponding probability for all particles,
\begin{equation}
   \frac{P(\mathrm{break}\mid\mathrm{NA})}{P(\mathrm{break})}.
\end{equation}
We divide the particles into deciles according to the magnitude of their nonaffine motion. As shown in Fig.~\ref{fig:na-percentile-bond-breaking}, particles with the largest nonaffine displacements are significantly more likely to change neighbors, whereas particles in the lowest 70\% have enrichment factors below unity and are less likely to do so. This observation agrees with experiments by Chikkadi and Schall~\cite{ChikkadiSchall2012}, who used bond breaking to identify irreversible rearrangements associated with nonaffine motion. Figure~\ref{fig:na-percentile-bond-breaking} also shows no systematic dependence on the polydispersity $\delta$; the results are similar across all five data sets.

\section{Locally defined nonaffine motion: $D^2_{\text{min}}$}

\subsection{Defining locality through neighbor selection}

Introduced by Falk and Langer in the context of shear transformation zones, $D^2_{\min}$ measures local nonaffine motion~\cite{FalkLanger1998}. It has since become one of the most widely used quantities for identifying plastic rearrangements in amorphous materials~\cite{MaloneyLemaitre2006,NicolasRottler2018,FalkLanger2011Review,cubuk15}. The idea is straightforward: the relative motion of a particle's neighbors is fitted to a single affine transformation, and $D^2_{\min}$ is the residual of that fit.

For a particle $i$, let $N_i$ denote its set of neighbors, and let $\mathbf{r}_{ij}$ be the position of neighbor $j\in N_i$ relative to particle $i$. Over a time interval $\Delta t$, or equivalently a strain interval $\Delta\gamma$, we determine the affine transformation matrix $\mathbf{E}$ that best maps $\mathbf{r}_{ij}(t)$ to $\mathbf{r}_{ij}(t+\Delta t)$. Thus,
\begin{equation}
D^2_{\min}(i,t)
=
\frac{1}{|N_i|}
\min_{\mathbf{E}}
\sum_{j\in N_i}
\left\|
\mathbf{r}_{ij}(t+\Delta t)
-
\mathbf{E}\,\mathbf{r}_{ij}(t)
\right\|^2.
\end{equation}

The quantity $D^2_{\min}$ is therefore the least-squares fit error and quantifies how strongly the motion around particle $i$ deviates from a collective affine deformation. The factor $1/|N_i|$ normalizes the result by the number of neighbors, giving the mean squared residual per neighbor. Because rearrangements produce localized nonaffine motion, they generally correspond to large values of $D^2_{\min}$~\cite{NicolasRottler2018,cubuk15}.

The strength of this method is that, unlike the mean-flow measure in Sec.~\ref{sec:nonaffine}, it defines affine motion locally in both space and time. A large group of particles may move coherently in a way that differs from the mean flow, producing large mean-flow nonaffine displacements without bond breaking within the group. In this case, the locally fitted transformation $\mathbf{E}$ may differ from the time-averaged mean flow, yet $D^2_{\min}$ remains small because the relative particle motions are consistent with that transformation. By contrast, a large value of $D^2_{\min}$ identifies a region in which no single affine transformation adequately describes the relative motion, thereby capturing the local character of a rearrangement.

For the present work, the important point is that $D^2_{\min}$ depends explicitly on the chosen neighbor set $N_i$. We therefore consider three methods for defining neighborhoods in highly polydisperse systems: (a) radical Delaunay triangulation (RDT), (b) a fixed number $K$ of nearest neighbors (KNN), and (c) a pairwise cutoff distance (PCD).

To use RDT for $D^2_{\min}$, we extend the immediate-neighbor definition introduced in Sec.~\ref{sec:bondbreak} to include multiple topological rings. The direct RDT neighbors of particle $i$ form ring $\ell=1$. Ring $\ell=2$ contains particles whose shortest path to particle $i$ consists of two RDT edges, and subsequent rings are defined similarly. Using cumulative rings allows us to vary the neighborhood size and examine its influence on $D^2_{\min}$. RDT is size-aware and therefore well suited to highly polydisperse systems. It is also reciprocal: if particle $j$ lies within $\ell$ rings of particle $i$, then particle $i$ lies within $\ell$ rings of particle $j$. A drawback is that topological rings do not correspond to a uniform physical distance. In regions containing larger particles, the neighborhood may extend farther in some directions, so it need not be spatially centered on the particle at its topological center.

\begin{figure*}[t]
  \centering
  \includegraphics[width=\textwidth]{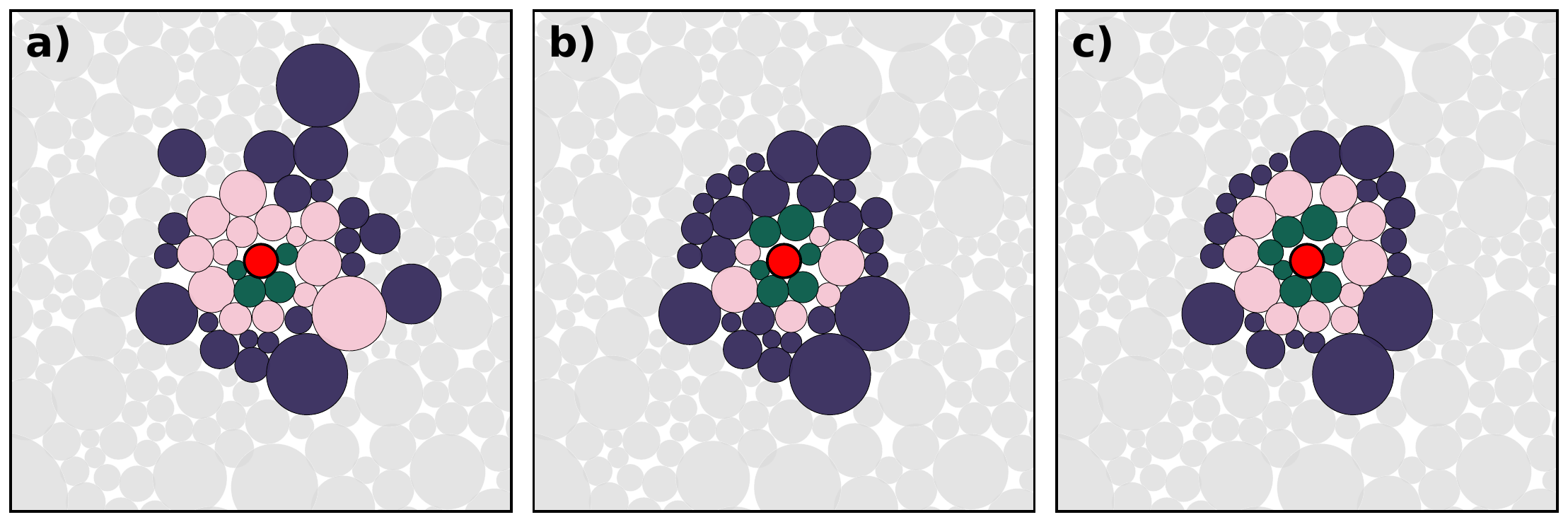}

  \caption{Growth of neighborhoods defined by the three methods in the $\delta=0.501$ simulation, shown for a representative central particle with $R_i/\langle R\rangle=0.91$. The central particle is red, and neighbors added as each tuning parameter increases are shown in successive color groups: (a) for PCD, increasing the cutoff from $C=1.0$ to $2.0$ to $3.3$ gives 4, 18, and 39 neighbors, respectively. The isolated large neighbor at the upper left illustrates that a large particle can satisfy the PCD criterion at a greater center-to-center separation than a small particle; (b) for KNN, the neighborhoods contain the first $K=6$, 12, and 39 particles ranked by surface-to-surface gap; (c) for RDT, cumulative topological rings $\ell=1$, 2, and 3 contain 7, 19, and 39 neighbors, respectively. The largest PCD and KNN neighborhoods were chosen to match the number of particles in the $\ell=3$ RDT neighborhood, illustrating that similar neighbor counts can correspond to different definitions of locality.}
  \label{fig:candidate-c-growth}
\end{figure*}
The other two methods select neighbors using distance. In a polydisperse system, however, one must first decide which distance is physically relevant. If neighbors are interpreted as particles that are touching or nearly touching, the surface-to-surface gap is more appropriate than the center-to-center distance. We therefore define
\begin{equation}
h_{ij}=r_{ij}-(R_i+R_j),
\end{equation}
where $r_{ij}$ is the center-to-center distance between particles $i$ and $j$, and $R_i$ and $R_j$ are their radii. This convention has been used in prior work~\cite{clararahola15,JiangSussmanWeeks2023,IllingWeeks2025NonaffinePolydisperse}. In the fixed-number KNN construction, the $K$ particles with the smallest gap distances are included in the neighborhood. Jiang \textit{et al.} noted that, for these simulation data, $K=15$ includes at least one complete neighbor shell around larger particles and at least two shells around smaller particles~\cite{JiangSussmanWeeks2023}. The KNN method is easy to tune by varying $K$, and ranking particles by their surface-to-surface gaps makes the method size-aware. One drawback is that reciprocity is not guaranteed: particle $j$ may be among the $K$ nearest neighbors of particle $i$ even when particle $i$ is not among the $K$ nearest neighbors of particle $j$.

The third method uses a pairwise cutoff distance (PCD), for which particles $i$ and $j$ are defined as neighbors whenever
\begin{equation}
r_{ij} \le C\,(R_i + R_j),
\label{eq:pcd}
\end{equation}
where $r_{ij}=\lVert\mathbf{r}_j-\mathbf{r}_i\rVert$ is the center-to-center distance, $R_i$ and $R_j$ are the particle radii, and $C$ is a dimensionless cutoff parameter. This method has been used by Yamamoto and Onuki~\cite{YamamotoOnuki1998}. The definition is particularly appealing for polydisperse systems because it is explicitly size-aware. For $C=1$, neighboring particles must touch or overlap, whereas $C>1$ includes particles separated by a gap and therefore defines larger neighborhoods. Reciprocity follows automatically from the symmetry under exchange of $i$ and $j$, and increasing $C$ expands the neighborhood without favoring a particular direction.

The three methods differ not only in their tuning parameters---$C$ for PCD, $K$ for KNN, and ring number $\ell$ for RDT---but also in how their neighborhoods expand as these parameters increase. Figure~\ref{fig:candidate-c-growth} illustrates these differences for a representative particle highlighted in red. When their parameters are chosen to produce neighborhoods of comparable size, the methods identify broadly similar groups of nearby particles, but the detailed shapes and compositions of those neighborhoods remain distinct.

\subsection{Characteristics of neighborhoods}

We reiterate the goal of defining neighbors: we seek a region around particle $i$ in which a single affine matrix $\mathbf{E}$ describes the local motion when particles are not rearranging, but for which $D^2_{\text{min}}$ becomes large when a rearrangement occurs~\cite{cubuk15}. Although $D^2_{\text{min}}$ is directly calculable, its precise numerical value is less important than its ability to identify regions of unusually nonaffine motion. There is therefore no universally correct way to define the neighbors. Instead, we examine the implications of different choices and provide practical recommendations for highly polydisperse samples. In this subsection, we compare the sizes and shapes of the neighborhoods defined by the three methods introduced above.

\begin{figure*}[t]
  \centering
  \includegraphics[width=\textwidth]{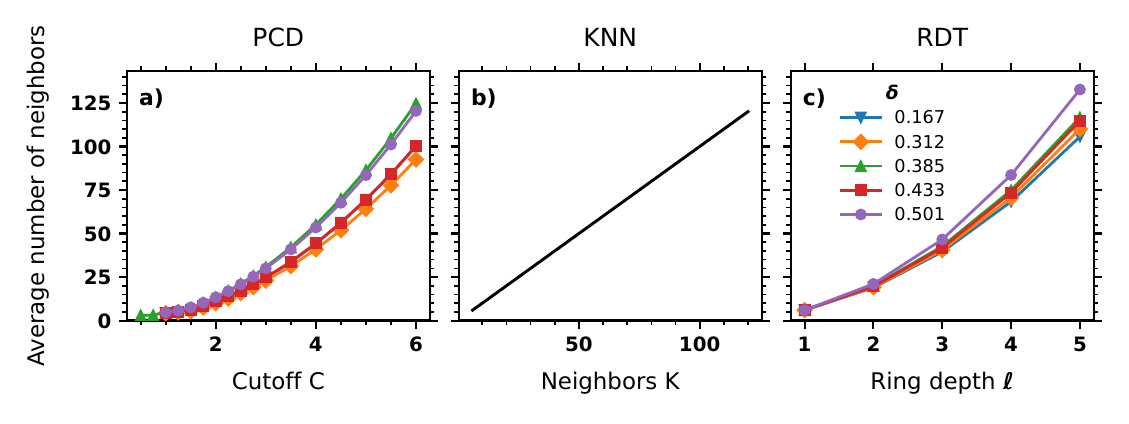} 
  \caption{Mean neighborhood size as a function of the tuning parameter for (a) pairwise cutoff distance (PCD), (b) fixed-number nearest neighbors (KNN), and (c) radical Delaunay triangulation (RDT). The curves represent the different polydispersities identified in the legend in (c). In (a), the blue and red curves nearly overlap. In (b), the number of neighbors equals $K$ by definition and therefore follows the line $y=x$.}
  \label{fig:neighbor-methods}
\end{figure*}

As shown in Fig.~\ref{fig:neighbor-methods}, each method allows the typical neighborhood size to be tuned. KNN sets the number of neighbors directly through $K$. For PCD and RDT, the mean number of neighbors is generally larger in more polydisperse samples at fixed $C$ or $\ell$. Although large particles constitute only a fraction of each sample, they tend to have more neighbors under both PCD and RDT, thereby increasing the mean neighbor count.

\begin{figure}[t]
  \centering
  \includegraphics[width=\columnwidth]{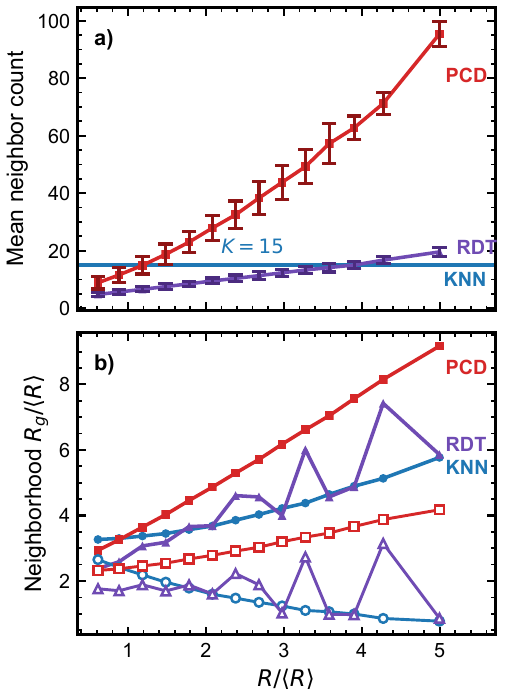}
  \caption{Particle-size dependence of the neighborhoods in the
highest-polydispersity simulation ($\delta=0.501$).
(a) Mean neighbor count as a function of the central-particle radius
$R/\langle R\rangle$. Error bars show one standard deviation of the
particle-level neighbor-count observations within each radius bin, pooled over the analyzed frames. For KNN, the neighbor count is fixed at $K=15$ and therefore has zero standard deviation.
(b) Neighborhood radius of gyration $R_g/\langle R\rangle$ (filled symbols) and neighborhood-shell thickness $(R_g-R)/\langle R\rangle$ (open symbols) as functions of $R/\langle R\rangle$. Results are shown for PCD with $C=2$, KNN with $K=15$, and first-ring RDT.}
  \label{fig:exp10-neighbors}
\end{figure}

\begin{figure}[t]
  \centering
  \includegraphics[width=\columnwidth]{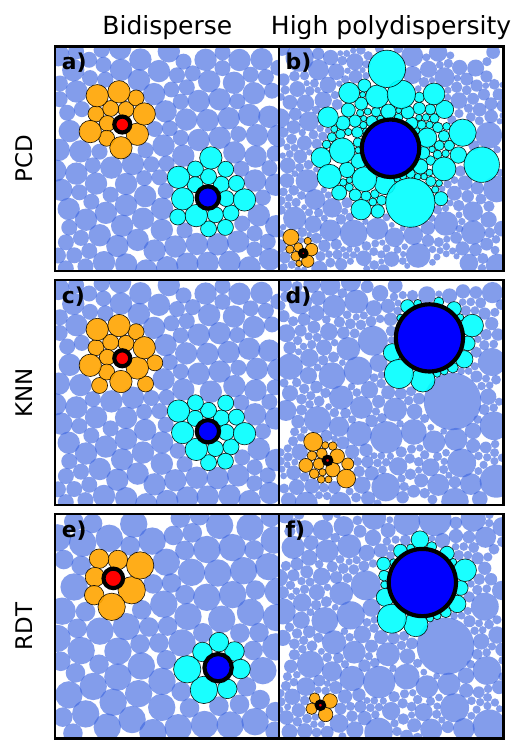}
  \caption{Comparison of neighborhoods selected around the same target particles using PCD with $C=2$, KNN with $K=15$, and first-ring RDT\@. The top row shows the bidisperse system, $\delta=0.167$, and the bottom row shows the highly polydisperse system, $\delta=0.501$. The columns correspond to PCD, KNN, and RDT, respectively. The small and large target particles are shown in red and blue, with their selected neighbors shown in orange and cyan. The three methods produce similar neighborhoods in the bidisperse system but noticeably different neighborhoods in the highly polydisperse system, particularly around the large particle.}
  \label{fig:same-particles-pcd-knn-rdt}
\end{figure}

To emphasize this size dependence, Fig.~\ref{fig:exp10-neighbors}(a) shows the mean number of neighbors as a function of particle radius $R$. For KNN, this number is constant by construction, with $K=15$ in this example. For PCD and RDT, larger particles have more neighbors, and the increase is especially pronounced for PCD\@. The first RDT ring provides a near-minimal topological shell around each particle; even so, the largest particles have nearly 20 neighbors on average. By contrast, fixing the KNN count at 15 may leave the largest particles incompletely surrounded, as illustrated in Fig.~\ref{fig:same-particles-pcd-knn-rdt}(e). Thus, when KNN is used for highly polydisperse samples, the choice of $K$ must account for the full particle-size distribution. PCD and RDT adapt the neighbor count to the size and local environment of the central particle and therefore more naturally surround particles across the size distribution. For PCD, however, the cutoff must be sufficiently large: values of $C\leq1$ generally select too few neighbors, as suggested by Fig.~\ref{fig:neighbor-methods}(a).

We next consider the spatial extent of each neighborhood by calculating its radius of gyration, $R_g$, from the distances $r_{ij}$ between particle $i$ and its neighbors. Figure~\ref{fig:exp10-neighbors}(b) shows the mean $R_g/\langle R\rangle$ as a function of particle radius for the most polydisperse sample; the other samples exhibit similar behavior. Larger particles generally have larger neighborhoods. For the parameter choices shown, the neighborhood dimensions are comparable to the nonaffine-motion correlation length, $\xi/\langle R\rangle\approx6$, discussed in Sec.~\ref{sec:nonaffine}. This comparison is most direct for PCD\@.

We also characterize the effective thickness of the neighborhood shell using $R_g-R$, shown by the open symbols in Fig.~\ref{fig:exp10-neighbors}(b). This quantity decreases with $R$ for KNN, remains approximately constant apart from fluctuations for RDT, and increases with $R$ for PCD\@. These trends are also visible in Fig.~\ref{fig:same-particles-pcd-knn-rdt}. For example, under KNN, a set of 15 neighbors forms a relatively thick shell around a small particle but only a thin shell around a large particle, as seen in Fig.~\ref{fig:same-particles-pcd-knn-rdt}(e).

\subsection{Core--shell analysis}

Figure~\ref{fig:exp10-nonaffine-motion} provides intuition for how neighborhood size may affect $D^2_{\text{min}}$. As discussed in Sec.~\ref{sec:nonaffine}, the nonaffine-motion field has a correlation length of $\xi/\langle R\rangle\sim5$. It is therefore reasonable to expect that a useful neighborhood size should be comparable to $\xi$. A much smaller neighborhood may contain too few particles to constrain the affine fit $\mathbf{E}$ reliably; for example, a fit based on only two or three neighbors has little redundancy. By contrast, a neighborhood much larger than $\xi$ combines subregions with different local motions. In that case, a large value of $D^2_{\text{min}}$ may reflect spatial variation of the flow within the neighborhood rather than a localized rearrangement.

In the limit where the neighborhood includes the entire system, $\mathbf{E}$ approaches the global affine deformation. For the simulations considered here, which have a uniform imposed strain rate, $D^2_{\text{min}}$ would then lose its local meaning. For a system with a spatially varying flow, a single system-wide matrix $\mathbf{E}$ would provide an even poorer description. These considerations suggest that there is a useful range of neighborhood sizes: the neighborhood should be large enough to constrain the affine fit but small enough to remain local. Ideally, the scientific conclusions drawn from $D^2_{\text{min}}$ should also be robust within this range. This issue is especially important in highly polydisperse systems, where the broad particle-size distribution introduces multiple relevant length scales.

To examine how neighborhood size influences the affine fit $\mathbf{E}$ and the nonaffine residual $D^2_{\text{min}}$, we introduce a core--shell analysis. For each central particle, we define two nested neighborhoods. The smaller neighborhood forms the core, while the shell consists of the particles that belong to the larger neighborhood but not to the core. We fit $\mathbf{E}$ using the core particles and then compare the fitting residuals in the core and shell. Each residual is normalized by the number of particles in the corresponding set, so it represents a mean squared fitting error per particle.

For example, consider the pairwise cutoff distance (PCD) method. For each particle $i$, we define a core neighborhood $N_{i,\mathrm{core}}$ using the cutoff $C$ and fit an affine matrix $\mathbf{E}_{\mathrm{core}}$ to the motion within that core. We then ask how well this matrix describes the additional particles included when the cutoff is increased from $C$ to $C+\Delta C$. These added particles form the shell, denoted $N_{i,\mathrm{shell}}$.

The residual corresponding to the core fit, $D^2_{\mathrm{core}}$, is given by
{\small
\begin{equation}
D^2_{\mathrm{core}}(i)
=
\frac{1}{|N_{i,{\mathrm{core}}}|}
\sum_{j\in N_{i,{\mathrm{core}}}}
\left\|
\mathbf{r}_{ij}(t+\Delta t)
-
\mathbf{E}_{\mathrm{core}}(i)\mathbf{r}_{ij}(t)
\right\|^2.
\end{equation}
}
We then ask how well $\mathbf{E}_{\mathrm{core}}(i)$ describes the motion of particles in the shell by calculating the residual

{\small
\begin{equation}
\begin{aligned}
D^2_{\mathrm{shell}\leftarrow\mathrm{core}}(i,C)
&= \frac{1}{|N_{i,{\mathrm{shell}}}|} \times \\
&\sum_{j\in N_{i,{\mathrm{shell}}}}
  \left\| \mathbf{r}_{ij}(t+\Delta t)
- \mathbf{E}_{\mathrm{core}}(i)\mathbf{r}_{ij}(t) \right\|^2.
\end{aligned}
\end{equation}
}

Finally, we evaluate the ratio of the shell and core residuals for each value of $C$:
\begin{equation}
Q(C)
=
\frac{\mathrm{median}_{i,t}\left[D^2_{\mathrm{shell}\leftarrow\mathrm{core}}(i,C)\right]}
{\mathrm{median}_{i,t}\left[D^2_{\mathrm{core}}(i,C)\right]}.
\end{equation}
Because $\mathbf{E}_{\mathrm{core}}$ is optimized for the core rather than the shell, we expect $Q$ to exceed unity on average, although this is not guaranteed for every particle and time. A value of $Q(C)>1$ means that the affine matrix fitted to the core describes the newly added shell particles less accurately than it describes the core particles. The medians are taken over all particles $i$ and initial times $t$.

\begin{figure*}[t]
  \centering
  \includegraphics[width=1\textwidth]{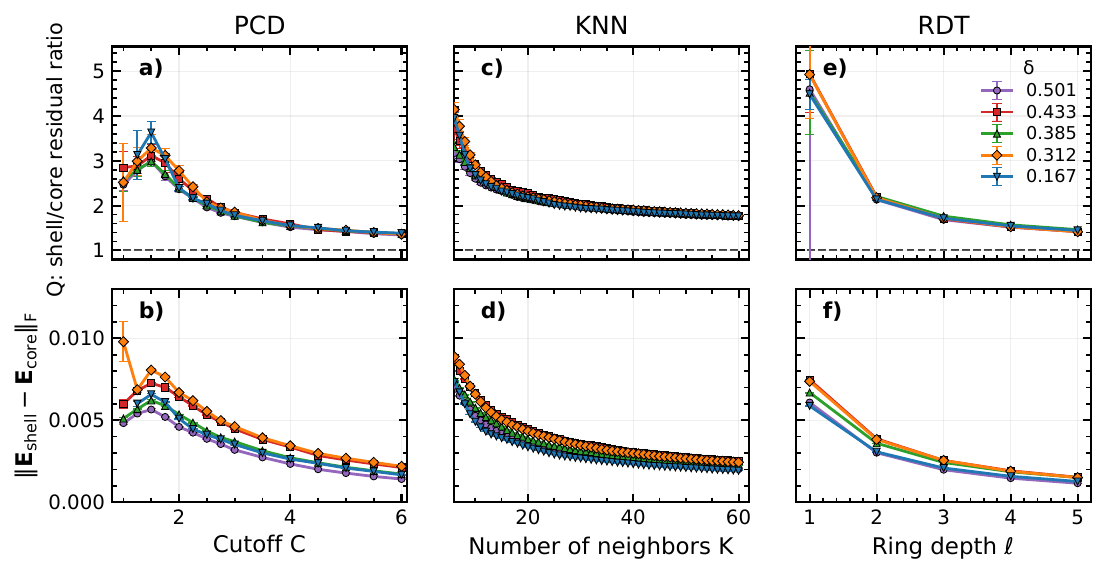}
  \caption{Core-shell comparison of affine fits for three methods of defining neighbors. Each column of panels corresponds to one method. The first column (panels a,b) shows the pairwise cutoff distance (PCD) results, the second column (panels c,d) shows the fixed-number nearest-neighbor results (KNN), and the third column (panels e,f) shows the radical Delaunay triangulation (RDT) results. The top row displays the ratio $Q$, defined as the ratio of the residual obtained by fitting the shell using $\mathbf{E}_{\mathrm{core}}$ to the residual of the core. The dashed line marks $Q=1$, where the two residuals are equal. The bottom row shows the difference between $\mathbf{E}_{\mathrm{core}}$ and $\mathbf{E}_{\mathrm{shell}}$ measured using the Frobenius norm. Error bars indicate the standard error of the mean across analyzed time frames. For PCD, the core is defined by the cutoff \(C\), and the shell contains particles added when the cutoff is increased from \(C\) to \(C+\Delta C\), with \(\Delta C=0.5\). For KNN, the core is defined as the \(K\) nearest neighbors and the shell as neighbors ranked \(K+1\) to \(2K\). For RDT, the core is given by the ring depth indicated, and the shell is the next ring outward.}
  \label{fig:core-vs-shell}
\end{figure*}

The results for $Q$ are shown in the top row of Fig.~\ref{fig:core-vs-shell}, which plots $Q$ as a function of the parameter controlling the core-neighborhood size for the three neighbor definitions. We find $Q>1$ throughout. For all three methods, $Q$ decreases as the core neighborhood grows. A larger core samples more independently moving regions, so its fitted matrix approaches the global affine deformation imposed in the simulation. The core matrix then describes the core and shell with increasingly similar accuracy. Although the residuals in both regions increase as local distinctions are averaged out, they approach a common value, causing $Q$ to decrease toward unity. For smaller neighborhoods, the locally preferred affine transformations of the core and shell differ more strongly, so applying $\mathbf{E}_{\mathrm{core}}$ to the shell produces a larger value of $Q$.

To test this interpretation, we independently determine $\mathbf{E}_{\mathrm{shell}}(i)$ by fitting an affine transformation to the particles in $N_{i,{\mathrm{shell}}}$. We then quantify the difference between the core and shell matrices using the Frobenius norm,
\begin{equation}
\Delta E(C)
=
\mathrm{median}_{i,t}
\left[
\left\lVert
\mathbf{E}_{\mathrm{shell}}(i)
-
\mathbf{E}_{\mathrm{core}}(i)
\right\rVert_F
\right].
\end{equation}
The median is again taken over all particles $i$ and initial times $t$. The results, shown in the bottom row of Fig.~\ref{fig:core-vs-shell}, follow a trend similar to that of $Q$ in the top row. As the core neighborhood grows, the affine matrices fitted to the core and shell approach one another, as well as the global imposed deformation, and the Frobenius norm of their difference decreases. Both regions increasingly average over the heterogeneous motion, thereby washing out the local distinctions that $D^2_{\text{min}}$ is intended to detect.

For all the data shown in Fig.~\ref{fig:core-vs-shell}, the magnitudes of $Q$ and $\Delta E$ depend on the shell thickness; a thicker shell generally increases the contrast between the core and shell. Direct comparisons of magnitude among the three neighborhood methods are therefore not meaningful. Instead, the useful information is how these quantities change as the core neighborhood size is varied.

Intriguingly, the PCD results shown in Fig.~\ref{fig:core-vs-shell}(a,b) exhibit clear maxima near $C=1.5$--$1.8$ for all polydispersities, corresponding to roughly 6--8 neighbors, as shown in Fig.~\ref{fig:neighbor-methods}. This suggests that $C\approx1.6$ identifies a characteristic length scale at which the contrast between the core and shell is greatest. The robustness of this range across polydispersities strengthens the evidence for such a crossover scale. These neighborhoods are smaller than the nonaffine correlation length $\xi$. Because $\xi$ characterizes how rapidly nonaffine motion changes with position, it is reasonable that the $D^2_{\min}$ procedure selects a smaller region over which a single affine fit works well. Thus, the maxima near $C\approx1.6$ identify a particularly useful local scale for calculating $D^2_{\text{min}}$. The maxima also reflect the decrease in both quantities at small $C$, where the neighborhood contains too few particles. In this limit, the core and shell affine fits are similar, indicating that both sets sample comparable local regions. It remains unclear why KNN and RDT do not exhibit similar downturns at their smallest neighborhood sizes and therefore lack corresponding local maxima.

\subsection{Validation with nonaffine displacements}

In the previous section, we examined how the spatial behavior of $D^2_{\min}$ depends on the neighborhood definition. We now compare $D^2_{\min}$ with rearrangements identified using nonaffine motion (NA). Because the NA value of each particle does not require a neighbor definition, it provides a useful benchmark for evaluating how well the different neighborhoods capture rearrangement activity.

\begin{figure*}[t]
  \centering
  \includegraphics[width=\textwidth]{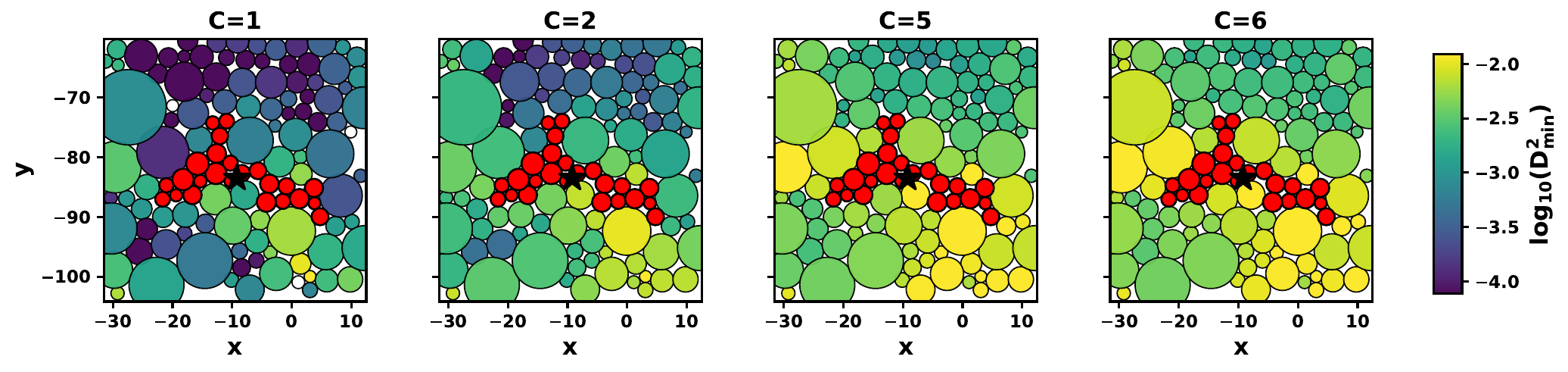}

  \caption{Visualization of $D^2_{\min}$ near a cluster of particles with high nonaffine motion. Each panel shows $\log_{10}(D^2_{\min})$ calculated using PCD with a different cutoff $C$. The red particles form a cluster selected from the top 10\% of the NA distribution, and the black star marks the cluster's center of mass. Darker and lighter colors indicate lower and higher values of $D^2_{\min}$, respectively. Increasing $C$ broadens the high-$D^2_{\min}$ region around the cluster, illustrating how larger neighborhoods reduce the locality of the rearrangement signal. The spatial pattern changes little between $C=5$ and $C=6$, indicating that this broadening eventually plateaus.}
  \label{fig:d2min-validation}
\end{figure*}

We first identify particles in the top 10\% of the NA distribution. As shown in Fig.~\ref{fig:na-percentile-bond-breaking}, these particles are also more likely to change neighbors and therefore to be rearranging. We then restrict attention to spatially adjacent groups of at least three high-NA particles, which we refer to as clusters. This criterion emphasizes the local and collective character of rearrangements rather than isolated particles with unusually large NA motion. One such cluster is shown in red in Fig.~\ref{fig:d2min-validation}.

Having defined the high-NA clusters, we examine the behavior of $D^2_{\min}$ around each one. Figure~\ref{fig:d2min-validation} shows that $D^2_{\min}$ is elevated near a high-NA cluster and that its spatial extent depends on the neighborhood definition. For each cluster, we compute its center of mass and construct the radial mean profile $\langle D^2_{\min}(r)\rangle$ as a function of distance $r$ from that center. We define $\xi_1$ as the distance at which the profile has completed a fraction $1-1/e$ of its total decay toward the background value. A smaller $\xi_1$ indicates that the high-$D^2_{\min}$ signal remains localized near the NA-defined rearrangement, whereas a larger value indicates a more spatially extended signal.

\begin{figure*}[t]
  \centering
  \includegraphics[width=\textwidth]{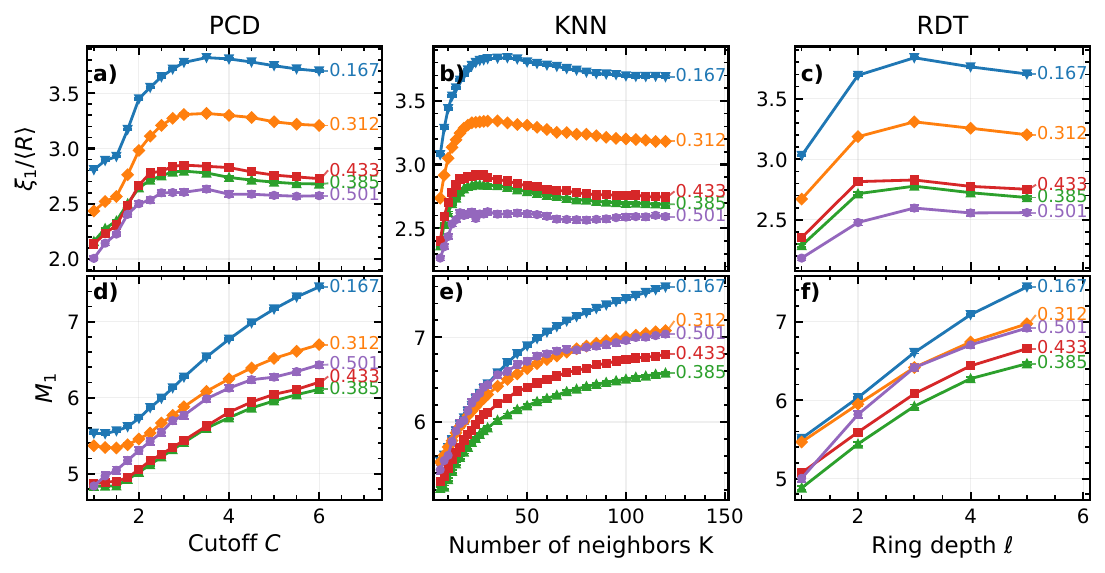}
  \caption{Comparison of locality and overlap enrichment for the three neighbor-definition methods. The columns show PCD, KNN, and RDT, and the curves represent the labeled polydispersities. The top row [panels (a)--(c)] shows the $1/e$ decay length $\xi_1$ of the radial $D^2_{\min}$ profile around NA-defined rearrangement clusters. The bottom row [panels (d)--(f)] shows the overlap enrichment $M_1$ between particles with large $D^2_{\min}$ and particles in high-NA clusters. Together, these quantities illustrate the tradeoff between agreement with the NA-based rearrangement measure and preservation of spatial locality as the neighborhood grows.}
  \label{fig:precision-locality-all-methods}
\end{figure*}

Figure~\ref{fig:precision-locality-all-methods} shows that $\xi_1$ is initially small for narrow neighborhoods, grows rapidly at intermediate neighborhood sizes, and reaches a plateau for sufficiently large neighborhoods. This trend is similar across polydispersities, although $\xi_1$ generally decreases as the polydispersity $\delta$ increases, suggesting that greater size disparity confines the nonaffine response to a narrower region. We interpret the large-neighborhood plateau as an estimate of the characteristic event size, with $\xi_1/\langle R\rangle\sim2.5$--$3.8$. This range is smaller than the nonaffine-motion correlation length, $\xi/\langle R\rangle\sim4.6$--$6.6$, although $2\xi_1/\langle R\rangle\sim5.0$--$7.6$ may be interpreted as the diameter of the high-$D^2_{\min}$ region surrounding a high-NA cluster.

Next, we define an overlap metric $M_1$ as
\begin{equation}
M_1 =
\frac{P(i\in S_D \mid i \in S_{\mathrm{NA}})}{P(i\in S_D)},
\end{equation}
where $S_D$ is the set of particles in the top 10\% of the $D^2_{\min}$ distribution, and $S_{\mathrm{NA}}$ is the set of high-NA particles belonging to the clusters defined above. The denominator is $P(i\in S_D)=0.1$ by construction. Thus, $M_1>1$ indicates that membership in $S_{\mathrm{NA}}$ increases the likelihood that a particle also belongs to $S_D$. As shown in Fig.~\ref{fig:precision-locality-all-methods}(d)--(f), $D^2_{\min}$ and nonaffine motion are strongly associated, and their overlap increases with neighborhood size. This behavior is expected because the fitted matrix $\mathbf{E}$ approaches the globally imposed deformation as the neighborhood grows; consequently, $D^2_{\min}$ becomes increasingly similar to a local average of the squared nonaffine motion. The results are robust when the percentile thresholds are varied from 5\% to 20\%.

These observations suggest that a small neighborhood better distinguishes the local affine deformation from the globally imposed deformation. It can therefore highlight regions with strong internal rearrangement while distinguishing them from groups that move coherently relative to the mean flow. A larger neighborhood produces a stronger statistical association with NA by averaging over more particles, but at the cost of spatial locality. Increasing the neighborhood beyond the crossover evident in Fig.~\ref{fig:precision-locality-all-methods} provides little additional benefit because the neighborhood begins to average over distinct regions.

\subsection{Validation through bond breaking}

Finally, we wish to understand how the neighborhood size used for $D^2_{\min}$ relates to bond breaking.  For bond breaking, rearrangements manifest as changes in the local particle-connectivity network, whereas nonaffine motion and $D^2_{\min}$ both relate to particle displacements rather than topology.  As discussed in Sec.~\ref{sec:bondbreak}, neighbors for bond breaking are defined using the first ring of radical Delaunay neighbors.  Again, we define a metric comparing the particles with large $D^2_{\min}$ and particles that undergo neighbor changes:
\begin{equation}
M_2 =
\frac{P(i\in S_D \mid i \in S_{\mathrm{BB}})}{P(i\in S_D)},
\end{equation}
where $S_D$ is the set of particles in the top 10\% of $D^2_{\min}$ values, and $S_{\mathrm{BB}}$ is the set of all particles that change one or more neighbors between $t$ and $t+\Delta t$. The denominator is $P(i\in S_D)=0.1$ by construction. As with $M_1$, a value of $M_2>1$ indicates that particles that change neighbors are more likely to have large values of $D^2_{\text{min}}$.

\begin{figure*}[t]
  \centering
  \includegraphics[width=0.9\textwidth]{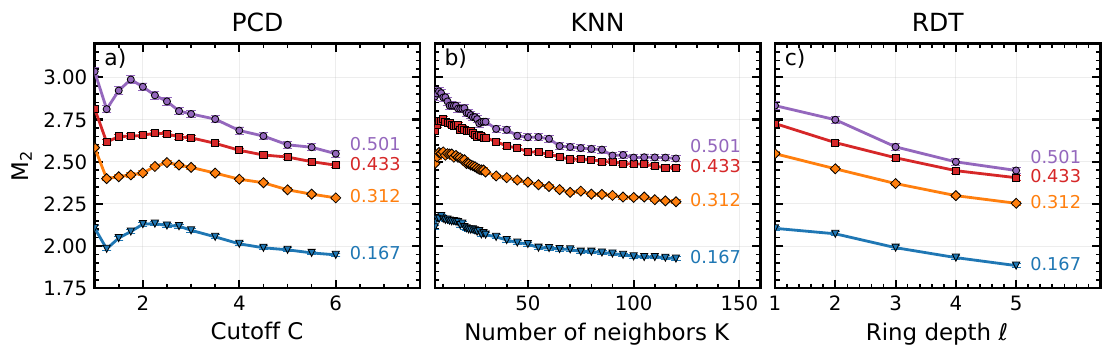}
  \caption{Metric~2 as a function of the neighbor-definition parameter for the three methods considered: pairwise cutoff distance (PCD), radical Delaunay triangulation (RDT), and fixed-number nearest neighbors (KNN). Across all polydispersities (as labeled in each panel), $M_2$ remains greater than 1, confirming that particles with large $D^2_{\min}$ are enriched in bond-breaking events. The strongest enrichment is obtained for highly local neighborhoods: for RDT the maximum occurs at $\ell=1$, for KNN it occurs for small neighborhoods corresponding to roughly 5--6 neighbors, and for PCD it occurs at $C=1$ for most polydispersities, with the bidisperse case ($\delta = 0.167$) peaking at the larger value $C=2.25$. Overall, PCD yields the largest $M_2$ values and therefore the strongest correspondence between large $D^2_{\min}$ and bond breaking.}
  \label{fig:metric2}
\end{figure*}

The results are presented in Fig.~\ref{fig:metric2}. First, across all neighbor definitions and polydispersities, we find that $M_2>1$. This confirms that particles experiencing bond breaking are more likely to also have large $D^2_{\min}$ values. Second, $M_2$ tends to be larger at higher polydispersities, which suggests that as size disparity increases, the association between bond breaking and large $D^2_{\min}$ values becomes more pronounced.

The behavior does, however, depend on the neighbor-definition method. In all cases, the data in Fig.~\ref{fig:metric2} show a maximum of $M_2$ for small neighborhood sizes. For RDT neighborhoods, the largest value of $M_2$ occurs at the first ring, $\ell=1$, across all polydispersities. This is sensible because these are the same neighbors used to define bond breaking. The result suggests that the bond-breaking signal is highly localized: including more neighbors weakens the correlation between large $D^2_{\min}$ values and changes in the local network. For KNN, the maximum $M_2$ occurs for neighborhoods containing roughly 5--6 particles across all polydispersities. For PCD, the maximum occurs within $C=1$--$2.25$, depending on polydispersity. However, the local maximum at $C=1$ likely arises because $D^2_{\min}$ is poorly constrained when the neighborhood contains only particles in direct contact, while bond breaking removes those contacts by definition. Large $D^2_{\min}$ is therefore correlated with bond breaking for a partly trivial reason. The secondary maxima near $C\approx1.75$--$2.50$ more likely identify meaningful neighborhood sizes for maximizing $M_2$. These PCD neighborhoods provide the largest values of $M_2$ in Fig.~\ref{fig:metric2}.

\subsection{Recommendation for neighborhood definition}

Overall, these results lead to a clear recommendation: for highly polydisperse systems such as those studied here, PCD should be used as the primary neighborhood definition for calculating $D^2_{\min}$ and detecting rearrangements with this measure. Recall that the goal is to define a neighborhood for which large values of $D^2_{\min}$ identify rearranging regions, while also ensuring that the definition performs consistently across samples with different polydispersities. Figures~\ref{fig:precision-locality-all-methods}(d--f) and~\ref{fig:metric2} show that every neighborhood method and neighborhood size considered produces some correlation between $D^2_{\min}$, nonaffine motion, and bond breaking. However, maximizing this correlation is not the only objective, since an excessively strong correlation may arise for trivial reasons. Instead, $D^2_{\min}$ should provide useful information that is complementary to the other measures of rearrangement.

We reject the KNN method because, at high polydispersity, using a fixed number of neighbors can leave large particles without complete angular coverage, as illustrated in Fig.~\ref{fig:same-particles-pcd-knn-rdt}(e). Although this problem can be reduced by increasing the number of neighbors, the appropriate value of $K$ would need to be tuned separately for each sample according to its particle-size distribution and largest particle size ratio.

In contrast, the PCD method has several desirable properties. The neighborhood criterion in Eq.~\ref{eq:pcd} depends on both particle radii and therefore adapts to each particle pair. The open squares in Fig.~\ref{fig:exp10-neighbors}(b) show that larger particles are surrounded by neighborhoods with a somewhat greater radial extent. This behavior is useful because larger particles should influence the surrounding flow over a larger region. More generally, one would like the spatial extent of the neighborhood to reflect the size of the central particle, as suggested by Fig.~\ref{fig:same-particles-pcd-knn-rdt}(a,d).

We recommend $C=2.0$ as the optimal choice because it balances the competing metrics considered here. Figure~\ref{fig:core-vs-shell} shows that $C\approx1.6$ maximizes the contrast between the core and shell. Figure~\ref{fig:metric2} shows that $C\approx1.75$--$2.5$ maximizes the correlation between bond breaking and $D^2_{\min}$. Figure~\ref{fig:precision-locality-all-methods}(a) shows that $C\approx2$ gives a decay length close to the asymptotic value reached for large neighborhoods, without requiring an unnecessarily large neighborhood. Finally, this choice gives approximately 15 neighbors on average, as shown in Fig.~\ref{fig:neighbor-methods}(a). This is similar to the number used in prior work, while allowing the number of neighbors to adapt to the size of the central particle [Fig.~\ref{fig:exp10-neighbors}(a)].

A comparison of correlation length scales further supports the choice of $C=2$. As described in Sec.~\ref{sec:nonaffine}, the nonaffine-motion field in the simulation data has a correlation length of $\xi_{\mathrm{NA}}/\langle R\rangle\approx4.6$--$6.6$. For PCD with $C=2$, the corresponding correlation length of the $D^2_{\min}$ field ranges from $\xi_{D^2}/\langle R\rangle\approx4.5$ to $5.5$ across the five polydispersities. These values are comparable to, although generally somewhat smaller than, the correlation lengths of the nonaffine-motion field. The cluster-centered analysis in Fig.~\ref{fig:precision-locality-all-methods}(a) provides a consistent geometric interpretation. At $C=2$, the diameter of the region with elevated $D^2_{\min}$ is $2\xi_1/\langle R\rangle\approx5.0$--$6.9$ across the five polydispersities. 

If a more localized measure of $D^2_{\min}$ is desired, $C=1.5$ provides a reasonable alternative. This choice enhances the core--shell contrast [Fig.~\ref{fig:core-vs-shell}(a,b)] and produces a $D^2_{\min}$ field that can vary more rapidly in space [Figs.~\ref{fig:d2min-validation} and~\ref{fig:precision-locality-all-methods}(a)]. It may therefore be preferable when the relevant rearrangements or deformation gradients vary over shorter spatial scales.

\section{Case study: experimental granular-flow data}
\label{sec:illing}

We next use the experimental data of Illing and Weeks~\cite{IllingWeeks2025NonaffinePolydisperse} to test our neighborhood-definition framework in a real, highly polydisperse granular flow. In their experiment, hard acrylic disks of different sizes are driven back and forth through a two-dimensional L-shaped channel by mechanical plungers, producing rearrangements within a spatially heterogeneous flow. Unlike the homogeneous shear used in the simulations of Jiang \emph{et al.}, the experimental flow has a velocity gradient that varies with position. The experiment therefore provides a useful test of the neighborhood definitions under more spatially complex flow conditions. To remain consistent with the original analysis, we treat the data as described by Illing and Weeks~\cite{IllingWeeks2025NonaffinePolydisperse}. In particular, we use $\Delta t=5~\mathrm{s}$ to measure particle displacements and exclude the initial transient behavior from the analysis.

\begin{figure*}[t]
  \centering
  \includegraphics[width=0.9\textwidth]{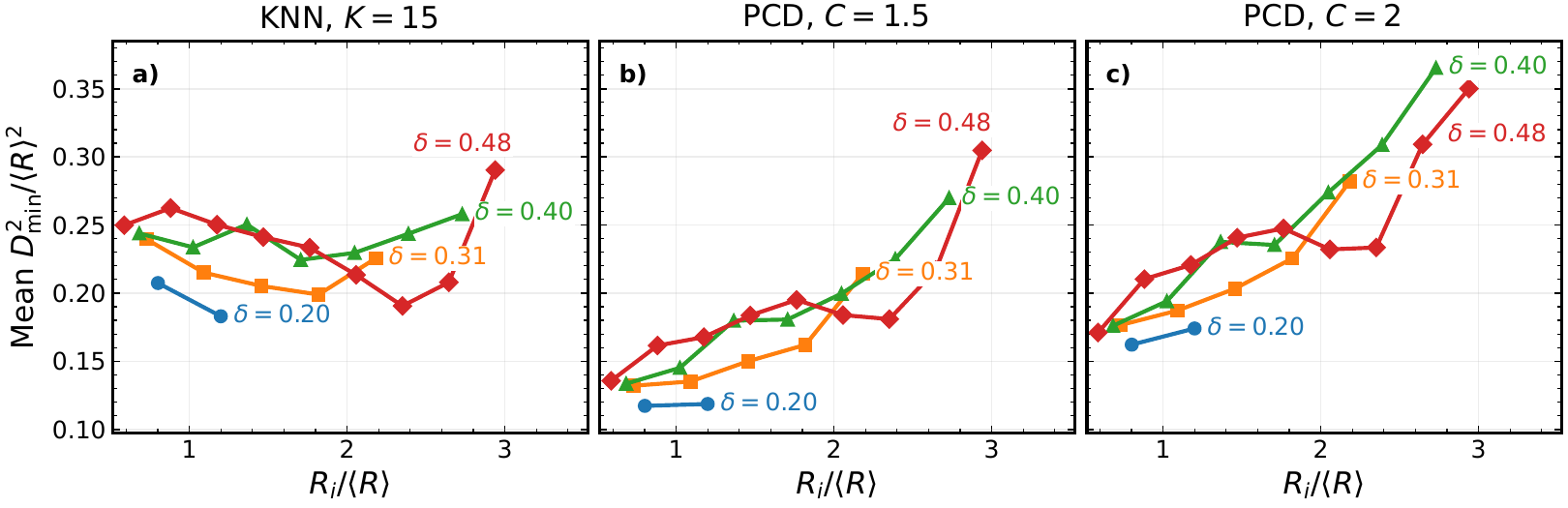}
  \caption{Particle-radius dependence of \(D^2_{\min}\) for two neighborhood definitions. The data are labeled by their polydispersities. The plotted quantity is the mean \(D^2_{\min}/\langle R\rangle^2\), where \(R\) is the particle radius.
  Panel (a) uses the fixed-neighbor definition KNN with \(K=15\), matching the convention used by Illing and Weeks.
  Panels (b) and (c) use the PCD definition with \(C=1.5\) and \(C=2\), respectively.
  The comparison shows that the size dependence of \(D^2_{\min}\) depends on the neighborhood used in the affine fit.}
  \label{fig:illing_d2_size}
\end{figure*}

Two questions considered by Illing and Weeks are how $D^2_{\min}$ depends on particle size within a given sample and how it depends on polydispersity across different samples. To address the first question, Fig.~\ref{fig:illing_d2_size} shows $D^2_{\min}/\langle R\rangle^2$ as a function of $R/\langle R\rangle$ for four samples. We use KNN to define the neighborhoods in panel (a) and PCD in panels (b) and (c). Each point represents an average of $D^2_{\min}$ over all particles with the same radius. Such averaging is possible because the samples are composed of disks with a discrete set of radii; one of the samples is bidisperse, with $\delta=0.20$.

Our KNN results do not reproduce the previously reported values exactly, likely because of differences in implementation details, but they agree qualitatively with the earlier work: $D^2_{\min}/\langle R\rangle^2$ decreases with increasing particle size~\cite{IllingWeeks2025NonaffinePolydisperse}. In contrast, when the neighborhoods are defined using PCD, $D^2_{\min}/\langle R\rangle^2$ generally increases with particle size, as shown in Fig.~\ref{fig:illing_d2_size}(b,c). This comparison is not intended to establish that one trend is correct and the other is incorrect. Rather, it demonstrates that the apparent size dependence of $D^2_{\min}$ is conditional on the neighborhood definition. Illing and Weeks emphasized that $D^2_{\min}$ depends on the chosen neighbors; our comparison extends this observation by showing that, in a highly polydisperse system, changing the definition of locality can reverse the apparent dependence of $D^2_{\min}$ on particle size.

One possible explanation for the increase in $D^2_{\min}$ with particle size under PCD is that the experimental flow varies rapidly with position~\cite{IllingWeeks2025NonaffinePolydisperse}. With $C=2$, the physical extent of a PCD neighborhood increases with the size of the central particle, as illustrated in Fig.~\ref{fig:exp10-neighbors}(b) for the simulation data. For the largest particles, the neighborhood typically spans a region over which the local deformation field varies appreciably. A single affine transformation would then provide a poorer description of the motion throughout that neighborhood, resulting in a larger value of $D^2_{\min}$ even in the absence of a correspondingly stronger local rearrangement.  To check this, we compare the trends for $C=1.5$ and $C=2.0$ in Fig.~\ref{fig:illing_d2_size}(b,c); the increase with $R_i/\langle R\rangle$ is reduced for the smaller neighborhood.  For example, the bidisperse sample loses any dependence on particle size in Fig.~\ref{fig:illing_d2_size}(b).

\begin{figure}[t]
  \centering
  \includegraphics[width=0.8\columnwidth]{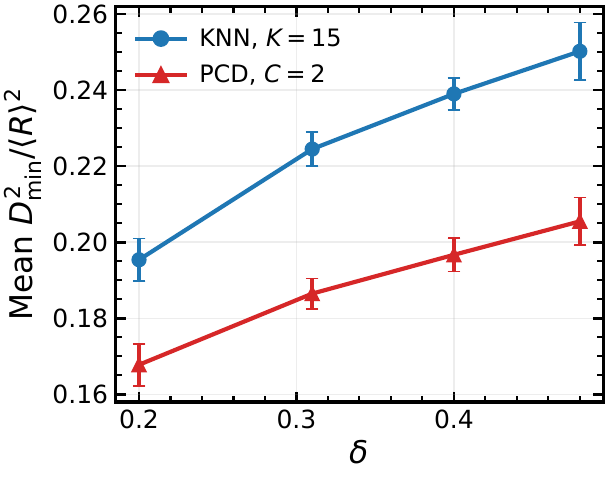}
  \caption{Mean \(D^2_{\min}/\langle R\rangle^2\) as a function of polydispersity for KNN with \(K=15\) and PCD with \(C=2\).
  Both definitions show an increase with polydispersity, indicating that the overall polydispersity trend is robust even though the particle size dependence depends on the neighborhood definition. Error bars indicate the standard error of the mean across experimental runs.
  }
  \label{fig:d2min-polydispersity-knn-pcd}
\end{figure}

Illing and Weeks reported that the normalized quantity \(D^2_{\min}/\langle R\rangle^2\) increases with polydispersity. Figure~\ref{fig:d2min-polydispersity-knn-pcd} shows the same qualitative increase for both KNN with \(K=15\) and PCD with \(C=2\), indicating that this global trend is robust to the neighborhood definition.

\section{Conclusion}

We have discussed three distinct methods for detecting rearrangements in highly polydisperse particulate materials.  Each has strengths and weaknesses.  Each correlates somewhat with the other two.  The three methods require varying amounts of adaptation to highly polydisperse materials, where particle sizes can vary by large factors within a sample.

The first method measures nonaffine motion, where affine motion is defined as motion varying linearly as a function of position, and the nonaffine motion is the component of an individual particle's displacement after subtracting the affine expectation.  The strength of this idea is that it is independent of the sample's particle-size distribution.  The weakness is that it relies on the affine motion being well-defined:  for example, it helps if the mean flow of the sample is time-independent.  Furthermore, this method could highlight a large group of particles that momentarily move collectively in a way distinct from the affine flow: while the group is moving in an unusual fashion, there may be no rearrangements occurring within it.

The second method considers when particles change their nearest neighbors (also termed bond breaking). While the literature offers multiple ways to define nearest neighbors, we recommend radical Delaunay triangulation for highly polydisperse samples. This method accounts for the radii of all particles when defining neighbors. The definition is symmetric: if particle $i$ is a neighbor of particle $j$, then particle $j$ is a neighbor of particle $i$. One drawback is that neighbors become poorly defined at the edges of a sample; in such cases, a distance cutoff may be needed~\cite{Weeks2000StructuralRelaxation}. Large particles also generally have more neighbors and are therefore more likely to change neighbors at any instant. Thus, the significance of bond breaking depends on context.

The final method uses the quantity $D^2_{\min}$ introduced by Falk and Langer~\cite{FalkLanger1998}.  This method requires defining a local neighborhood.  The displacements in that neighborhood are then fitted to an affine flow matrix:  that is, the matrix that best describes a flow that depends linearly on position.  The least-squares error for this flow matrix is the quantity $D^2_{\min}$.  The advantage of this method is that the description is local in space and time, making it well suited to time-varying flows.  A challenge is its dependence on the choice of neighbors, which becomes nontrivial for highly polydisperse systems.  Here we recommend using a pairwise cutoff distance, defining neighbors as those with centers separated by a distance less than $C$ times the sum of the two particles' radii~\cite{YamamotoOnuki1998}.  This definition is also symmetric (two particles are mutually neighbors).  Our analysis shows that using this method, with $C=2$, is optimal for capturing rearrangements.  Moreover, one can decrease $C$ if the flow varies strongly in space, to localize the meaning of $D^2_{\min}$.  However, a weakness of $D^2_{\text{min}}$ is that the results can depend qualitatively on the neighborhood choice, as illustrated by the case study in Sec.~\ref{sec:illing}.

Although our analysis focuses on two-dimensional systems, the framework
applies naturally in three dimensions. Both recommended neighborhood
definitions---radical Delaunay triangulation for bond breaking and the
pairwise cutoff for $D^2_{\min}$---work directly in either dimension.
By contrast, KNN requires a larger value of $K$ in three dimensions
because the number of particles needed to surround a central particle
scales with surface area rather than perimeter. The size-dependent
coverage discrepancy illustrated in
Fig.~\ref{fig:same-particles-pcd-knn-rdt}(e) should therefore be even
greater in three dimensions.

The framework is also not limited to driven systems. In quiescent
particle-resolved systems, the mean affine motion is often zero after
removing any global drift, while neighbor changes and $D^2_{\min}$ can
still identify thermally driven or aging-related rearrangements over an
appropriately chosen lag time.

We reiterate that the goals of all of these methods are to find ways to distinguish ``interesting'' particle motions from ``typical'' particle motions. These are subjective terms that depend on context, which is where the strengths and weaknesses of each method will matter.  Our recommendations are largely independent of the polydispersity.  In this way, future work can compare and contrast the behavior of different samples using analysis that translates smoothly between different systems with different particle-size distributions.

\begin{acknowledgments}
We thank Y.~Jiang for providing the simulation data used in this work, and P.~Illing for providing the experimental data. We also thank Y.~Jiang and D.~J.~Meer for helpful discussions. 
We acknowledge the use of OpenAI Codex and Prism as assistive tools for code editing, debugging, and proofreading. This material is based upon work supported by the National Science Foundation under Grant No. CBET-2306371.

\end{acknowledgments}

\hbadness=2000


\begin{thebibliography}{62}%
\makeatletter
\providecommand \@ifxundefined [1]{%
 \@ifx{#1\undefined}
}%
\providecommand \@ifnum [1]{%
 \ifnum #1\expandafter \@firstoftwo
 \else \expandafter \@secondoftwo
 \fi
}%
\providecommand \@ifx [1]{%
 \ifx #1\expandafter \@firstoftwo
 \else \expandafter \@secondoftwo
 \fi
}%
\providecommand \natexlab [1]{#1}%
\providecommand \enquote  [1]{``#1''}%
\providecommand \bibnamefont  [1]{#1}%
\providecommand \bibfnamefont [1]{#1}%
\providecommand \citenamefont [1]{#1}%
\providecommand \href@noop [0]{\@secondoftwo}%
\providecommand \href [0]{\begingroup \@sanitize@url \@href}%
\providecommand \@href[1]{\@@startlink{#1}\@@href}%
\providecommand \@@href[1]{\endgroup#1\@@endlink}%
\providecommand \@sanitize@url [0]{\catcode `\\12\catcode `\$12\catcode `\&12\catcode `\#12\catcode `\^12\catcode `\_12\catcode `\%12\relax}%
\providecommand \@@startlink[1]{}%
\providecommand \@@endlink[0]{}%
\providecommand \url  [0]{\begingroup\@sanitize@url \@url }%
\providecommand \@url [1]{\endgroup\@href {#1}{\urlprefix }}%
\providecommand \urlprefix  [0]{URL }%
\providecommand \Eprint [0]{\href }%
\providecommand \doibase [0]{https://doi.org/}%
\providecommand \selectlanguage [0]{\@gobble}%
\providecommand \bibinfo  [0]{\@secondoftwo}%
\providecommand \bibfield  [0]{\@secondoftwo}%
\providecommand \translation [1]{[#1]}%
\providecommand \BibitemOpen [0]{}%
\providecommand \bibitemStop [0]{}%
\providecommand \bibitemNoStop [0]{.\EOS\space}%
\providecommand \EOS [0]{\spacefactor3000\relax}%
\providecommand \BibitemShut  [1]{\csname bibitem#1\endcsname}%
\let\auto@bib@innerbib\@empty
\bibitem [{\citenamefont {Argon}(1979)}]{Argon1979}%
  \BibitemOpen
  \bibfield  {author} {\bibinfo {author} {\bibfnamefont {A.~S.}\ \bibnamefont {Argon}},\ }\bibfield  {title} {\bibinfo {title} {Plastic deformation in metallic glasses},\ }\href {https://doi.org/10.1016/0001-6160(79)90055-7} {\bibfield  {journal} {\bibinfo  {journal} {Acta Metallurgica}\ }\textbf {\bibinfo {volume} {27}},\ \bibinfo {pages} {47} (\bibinfo {year} {1979})}\BibitemShut {NoStop}%
\bibitem [{\citenamefont {Maloney}\ and\ \citenamefont {Lema{\^\i}tre}(2006)}]{MaloneyLemaitre2006}%
  \BibitemOpen
  \bibfield  {author} {\bibinfo {author} {\bibfnamefont {C.~E.}\ \bibnamefont {Maloney}}\ and\ \bibinfo {author} {\bibfnamefont {A.}~\bibnamefont {Lema{\^\i}tre}},\ }\bibfield  {title} {\bibinfo {title} {Amorphous systems in athermal, quasistatic shear},\ }\href {https://doi.org/10.1103/PhysRevE.74.016118} {\bibfield  {journal} {\bibinfo  {journal} {Phys. Rev. E}\ }\textbf {\bibinfo {volume} {74}},\ \bibinfo {pages} {016118} (\bibinfo {year} {2006})}\BibitemShut {NoStop}%
\bibitem [{\citenamefont {Bonn}\ \emph {et~al.}(2017)\citenamefont {Bonn}, \citenamefont {Denn}, \citenamefont {Berthier}, \citenamefont {Divoux},\ and\ \citenamefont {Manneville}}]{BonnEtAl2017YieldStress}%
  \BibitemOpen
  \bibfield  {author} {\bibinfo {author} {\bibfnamefont {D.}~\bibnamefont {Bonn}}, \bibinfo {author} {\bibfnamefont {M.~M.}\ \bibnamefont {Denn}}, \bibinfo {author} {\bibfnamefont {L.}~\bibnamefont {Berthier}}, \bibinfo {author} {\bibfnamefont {T.}~\bibnamefont {Divoux}},\ and\ \bibinfo {author} {\bibfnamefont {S.}~\bibnamefont {Manneville}},\ }\bibfield  {title} {\bibinfo {title} {Yield stress materials in soft condensed matter},\ }\href {https://doi.org/10.1103/RevModPhys.89.035005} {\bibfield  {journal} {\bibinfo  {journal} {Rev. Mod. Phys.}\ }\textbf {\bibinfo {volume} {89}},\ \bibinfo {pages} {035005} (\bibinfo {year} {2017})}\BibitemShut {NoStop}%
\bibitem [{\citenamefont {Barrat}\ and\ \citenamefont {Lema{\^\i}tre}(2011)}]{Barrat}%
  \BibitemOpen
  \bibfield  {author} {\bibinfo {author} {\bibfnamefont {J.-L.}\ \bibnamefont {Barrat}}\ and\ \bibinfo {author} {\bibfnamefont {A.}~\bibnamefont {Lema{\^\i}tre}},\ }\bibfield  {title} {\bibinfo {title} {Heterogeneities in amorphous systems under shear},\ }in\ \href {https://doi.org/10.1093/acprof:oso/9780199691470.003.0008} {\emph {\bibinfo {booktitle} {Dynamical Heterogeneities in Glasses, Colloids, and Granular Media}}},\ \bibinfo {editor} {edited by\ \bibinfo {editor} {\bibfnamefont {L.}~\bibnamefont {Berthier}}, \bibinfo {editor} {\bibfnamefont {G.}~\bibnamefont {Biroli}}, \bibinfo {editor} {\bibfnamefont {J.-P.}\ \bibnamefont {Bouchaud}}, \bibinfo {editor} {\bibfnamefont {L.}~\bibnamefont {Cipelletti}},\ and\ \bibinfo {editor} {\bibfnamefont {W.}~\bibnamefont {van Saarloos}}}\ (\bibinfo  {publisher} {Oxford University Press},\ \bibinfo {address} {Oxford},\ \bibinfo {year} {2011})\ pp.\ \bibinfo {pages} {264--297}\BibitemShut {NoStop}%
\bibitem [{\citenamefont {Tanguy}(2021)}]{Tanguy2021ElastoPlastic}%
  \BibitemOpen
  \bibfield  {author} {\bibinfo {author} {\bibfnamefont {A.}~\bibnamefont {Tanguy}},\ }\bibfield  {title} {\bibinfo {title} {Elasto-plastic behavior of amorphous materials: A brief review},\ }\href {https://doi.org/10.5802/crphys.49} {\bibfield  {journal} {\bibinfo  {journal} {Comptes Rendus Physique}\ }\textbf {\bibinfo {volume} {22}},\ \bibinfo {pages} {117} (\bibinfo {year} {2021})}\BibitemShut {NoStop}%
\bibitem [{\citenamefont {Desmond}\ \emph {et~al.}(2013)\citenamefont {Desmond}, \citenamefont {Young}, \citenamefont {Chen},\ and\ \citenamefont {Weeks}}]{Desmond2013ForcesEmulsionDroplets}%
  \BibitemOpen
  \bibfield  {author} {\bibinfo {author} {\bibfnamefont {K.~W.}\ \bibnamefont {Desmond}}, \bibinfo {author} {\bibfnamefont {P.~J.}\ \bibnamefont {Young}}, \bibinfo {author} {\bibfnamefont {D.}~\bibnamefont {Chen}},\ and\ \bibinfo {author} {\bibfnamefont {E.~R.}\ \bibnamefont {Weeks}},\ }\bibfield  {title} {\bibinfo {title} {Experimental study of forces between quasi-two-dimensional emulsion droplets near jamming},\ }\href {https://doi.org/10.1039/C3SM27287G} {\bibfield  {journal} {\bibinfo  {journal} {Soft Matter}\ }\textbf {\bibinfo {volume} {9}},\ \bibinfo {pages} {3424} (\bibinfo {year} {2013})}\BibitemShut {NoStop}%
\bibitem [{\citenamefont {Falk}\ and\ \citenamefont {Langer}(2011)}]{FalkLanger2011Review}%
  \BibitemOpen
  \bibfield  {author} {\bibinfo {author} {\bibfnamefont {M.~L.}\ \bibnamefont {Falk}}\ and\ \bibinfo {author} {\bibfnamefont {J.~S.}\ \bibnamefont {Langer}},\ }\bibfield  {title} {\bibinfo {title} {Deformation and failure of amorphous, solidlike materials},\ }\href {https://doi.org/10.1146/annurev-conmatphys-062910-140457} {\bibfield  {journal} {\bibinfo  {journal} {Annual Review of Condensed Matter Physics}\ }\textbf {\bibinfo {volume} {2}},\ \bibinfo {pages} {353} (\bibinfo {year} {2011})}\BibitemShut {NoStop}%
\bibitem [{\citenamefont {Ediger}(2000)}]{Ediger2000SpatiallyHeterogeneous}%
  \BibitemOpen
  \bibfield  {author} {\bibinfo {author} {\bibfnamefont {M.~D.}\ \bibnamefont {Ediger}},\ }\bibfield  {title} {\bibinfo {title} {Spatially heterogeneous dynamics in supercooled liquids},\ }\href {https://doi.org/10.1146/annurev.physchem.51.1.99} {\bibfield  {journal} {\bibinfo  {journal} {Annual Review of Physical Chemistry}\ }\textbf {\bibinfo {volume} {51}},\ \bibinfo {pages} {99} (\bibinfo {year} {2000})}\BibitemShut {NoStop}%
\bibitem [{\citenamefont {Sillescu}(1999)}]{Sillescu1999Heterogeneity}%
  \BibitemOpen
  \bibfield  {author} {\bibinfo {author} {\bibfnamefont {H.}~\bibnamefont {Sillescu}},\ }\bibfield  {title} {\bibinfo {title} {Heterogeneity at the glass transition: a review},\ }\href {https://doi.org/10.1016/S0022-3093(98)00831-X} {\bibfield  {journal} {\bibinfo  {journal} {Journal of Non-Crystalline Solids}\ }\textbf {\bibinfo {volume} {243}},\ \bibinfo {pages} {81} (\bibinfo {year} {1999})}\BibitemShut {NoStop}%
\bibitem [{\citenamefont {Weeks}\ \emph {et~al.}(2000)\citenamefont {Weeks}, \citenamefont {Crocker}, \citenamefont {Levitt}, \citenamefont {Schofield},\ and\ \citenamefont {Weitz}}]{Weeks2000StructuralRelaxation}%
  \BibitemOpen
  \bibfield  {author} {\bibinfo {author} {\bibfnamefont {E.~R.}\ \bibnamefont {Weeks}}, \bibinfo {author} {\bibfnamefont {J.~C.}\ \bibnamefont {Crocker}}, \bibinfo {author} {\bibfnamefont {A.~C.}\ \bibnamefont {Levitt}}, \bibinfo {author} {\bibfnamefont {A.}~\bibnamefont {Schofield}},\ and\ \bibinfo {author} {\bibfnamefont {D.~A.}\ \bibnamefont {Weitz}},\ }\bibfield  {title} {\bibinfo {title} {Three-dimensional direct imaging of structural relaxation near the colloidal glass transition},\ }\href {https://doi.org/10.1126/science.287.5453.627} {\bibfield  {journal} {\bibinfo  {journal} {Science}\ }\textbf {\bibinfo {volume} {287}},\ \bibinfo {pages} {627} (\bibinfo {year} {2000})}\BibitemShut {NoStop}%
\bibitem [{\citenamefont {Glotzer}(2000)}]{Glotzer2000HeterogeneousDynamics}%
  \BibitemOpen
  \bibfield  {author} {\bibinfo {author} {\bibfnamefont {S.~C.}\ \bibnamefont {Glotzer}},\ }\bibfield  {title} {\bibinfo {title} {Spatially heterogeneous dynamics in liquids: insights from simulation},\ }\href {https://doi.org/10.1016/S0022-3093(00)00225-8} {\bibfield  {journal} {\bibinfo  {journal} {Journal of Non-Crystalline Solids}\ }\textbf {\bibinfo {volume} {274}},\ \bibinfo {pages} {342} (\bibinfo {year} {2000})}\BibitemShut {NoStop}%
\bibitem [{\citenamefont {Berthier}\ and\ \citenamefont {Biroli}(2011)}]{BerthierBiroli2011GlassTransition}%
  \BibitemOpen
  \bibfield  {author} {\bibinfo {author} {\bibfnamefont {L.}~\bibnamefont {Berthier}}\ and\ \bibinfo {author} {\bibfnamefont {G.}~\bibnamefont {Biroli}},\ }\bibfield  {title} {\bibinfo {title} {Theoretical perspective on the glass transition and amorphous materials},\ }\href {https://doi.org/10.1103/RevModPhys.83.587} {\bibfield  {journal} {\bibinfo  {journal} {Reviews of Modern Physics}\ }\textbf {\bibinfo {volume} {83}},\ \bibinfo {pages} {587} (\bibinfo {year} {2011})}\BibitemShut {NoStop}%
\bibitem [{\citenamefont {Rabani}\ \emph {et~al.}(1997)\citenamefont {Rabani}, \citenamefont {Gezelter},\ and\ \citenamefont {Berne}}]{Rabani1997CageCorrelations}%
  \BibitemOpen
  \bibfield  {author} {\bibinfo {author} {\bibfnamefont {E.}~\bibnamefont {Rabani}}, \bibinfo {author} {\bibfnamefont {J.~D.}\ \bibnamefont {Gezelter}},\ and\ \bibinfo {author} {\bibfnamefont {B.~J.}\ \bibnamefont {Berne}},\ }\bibfield  {title} {\bibinfo {title} {Calculating the hopping rate for self-diffusion on rough potential energy surfaces: Cage correlations},\ }\href {https://doi.org/10.1063/1.475296} {\bibfield  {journal} {\bibinfo  {journal} {The Journal of Chemical Physics}\ }\textbf {\bibinfo {volume} {107}},\ \bibinfo {pages} {6867} (\bibinfo {year} {1997})}\BibitemShut {NoStop}%
\bibitem [{\citenamefont {Conrad}\ \emph {et~al.}(2006)\citenamefont {Conrad}, \citenamefont {Dhillon}, \citenamefont {Weeks}, \citenamefont {Reichman},\ and\ \citenamefont {Weitz}}]{Conrad2006SlowClusters}%
  \BibitemOpen
  \bibfield  {author} {\bibinfo {author} {\bibfnamefont {J.~C.}\ \bibnamefont {Conrad}}, \bibinfo {author} {\bibfnamefont {P.~P.}\ \bibnamefont {Dhillon}}, \bibinfo {author} {\bibfnamefont {E.~R.}\ \bibnamefont {Weeks}}, \bibinfo {author} {\bibfnamefont {D.~R.}\ \bibnamefont {Reichman}},\ and\ \bibinfo {author} {\bibfnamefont {D.~A.}\ \bibnamefont {Weitz}},\ }\bibfield  {title} {\bibinfo {title} {Contribution of slow clusters to the bulk elasticity near the colloidal glass transition},\ }\href {https://doi.org/10.1103/PhysRevLett.97.265701} {\bibfield  {journal} {\bibinfo  {journal} {Physical Review Letters}\ }\textbf {\bibinfo {volume} {97}},\ \bibinfo {pages} {265701} (\bibinfo {year} {2006})}\BibitemShut {NoStop}%
\bibitem [{\citenamefont {Yamamoto}\ and\ \citenamefont {Onuki}(1998)}]{YamamotoOnuki1998}%
  \BibitemOpen
  \bibfield  {author} {\bibinfo {author} {\bibfnamefont {R.}~\bibnamefont {Yamamoto}}\ and\ \bibinfo {author} {\bibfnamefont {A.}~\bibnamefont {Onuki}},\ }\bibfield  {title} {\bibinfo {title} {Dynamics of highly supercooled liquids: Heterogeneity, rheology, and diffusion},\ }\href {https://doi.org/10.1103/PhysRevE.58.3515} {\bibfield  {journal} {\bibinfo  {journal} {Phys. Rev. E}\ }\textbf {\bibinfo {volume} {58}},\ \bibinfo {pages} {3515} (\bibinfo {year} {1998})}\BibitemShut {NoStop}%
\bibitem [{\citenamefont {Shiba}\ \emph {et~al.}(2012)\citenamefont {Shiba}, \citenamefont {Kawasaki},\ and\ \citenamefont {Onuki}}]{shiba12}%
  \BibitemOpen
  \bibfield  {author} {\bibinfo {author} {\bibfnamefont {H.}~\bibnamefont {Shiba}}, \bibinfo {author} {\bibfnamefont {T.}~\bibnamefont {Kawasaki}},\ and\ \bibinfo {author} {\bibfnamefont {A.}~\bibnamefont {Onuki}},\ }\bibfield  {title} {\bibinfo {title} {Relationship between bond-breakage correlations and four-point correlations in heterogeneous glassy dynamics: {Configuration} changes and vibration modes},\ }\href {https://doi.org/10.1103/physreve.86.041504} {\bibfield  {journal} {\bibinfo  {journal} {Phys. Rev. E}\ }\textbf {\bibinfo {volume} {86}},\ \bibinfo {pages} {041504} (\bibinfo {year} {2012})}\BibitemShut {NoStop}%
\bibitem [{\citenamefont {Lundberg}\ \emph {et~al.}(2008)\citenamefont {Lundberg}, \citenamefont {Krishan}, \citenamefont {Xu}, \citenamefont {O'Hern},\ and\ \citenamefont {Dennin}}]{lundberg08}%
  \BibitemOpen
  \bibfield  {author} {\bibinfo {author} {\bibfnamefont {M.}~\bibnamefont {Lundberg}}, \bibinfo {author} {\bibfnamefont {K.}~\bibnamefont {Krishan}}, \bibinfo {author} {\bibfnamefont {N.}~\bibnamefont {Xu}}, \bibinfo {author} {\bibfnamefont {C.~S.}\ \bibnamefont {O'Hern}},\ and\ \bibinfo {author} {\bibfnamefont {M.}~\bibnamefont {Dennin}},\ }\bibfield  {title} {\bibinfo {title} {Reversible plastic events in amorphous materials},\ }\href {https://doi.org/10.1103/PhysRevE.77.041505} {\bibfield  {journal} {\bibinfo  {journal} {Physical Review E}\ }\textbf {\bibinfo {volume} {77}},\ \bibinfo {pages} {041505} (\bibinfo {year} {2008})}\BibitemShut {NoStop}%
\bibitem [{\citenamefont {Chen}\ \emph {et~al.}(2012)\citenamefont {Chen}, \citenamefont {Desmond},\ and\ \citenamefont {Weeks}}]{chen12}%
  \BibitemOpen
  \bibfield  {author} {\bibinfo {author} {\bibfnamefont {D.}~\bibnamefont {Chen}}, \bibinfo {author} {\bibfnamefont {K.~W.}\ \bibnamefont {Desmond}},\ and\ \bibinfo {author} {\bibfnamefont {E.~R.}\ \bibnamefont {Weeks}},\ }\bibfield  {title} {\bibinfo {title} {Topological rearrangements and stress fluctuations in quasi-two-dimensional hopper flow of emulsions},\ }\href {https://doi.org/10.1039/c2sm26023a} {\bibfield  {journal} {\bibinfo  {journal} {Soft Matter}\ }\textbf {\bibinfo {volume} {8}},\ \bibinfo {pages} {10486} (\bibinfo {year} {2012})},\ \Eprint {https://arxiv.org/abs/1105.3099} {arXiv:1105.3099} \BibitemShut {NoStop}%
\bibitem [{\citenamefont {Desmond}\ and\ \citenamefont {Weeks}(2015)}]{desmond15}%
  \BibitemOpen
  \bibfield  {author} {\bibinfo {author} {\bibfnamefont {K.~W.}\ \bibnamefont {Desmond}}\ and\ \bibinfo {author} {\bibfnamefont {E.~R.}\ \bibnamefont {Weeks}},\ }\bibfield  {title} {\bibinfo {title} {Measurement of stress redistribution in flowing emulsions},\ }\href {https://doi.org/10.1103/physrevlett.115.098302} {\bibfield  {journal} {\bibinfo  {journal} {Phys. Rev. Lett.}\ }\textbf {\bibinfo {volume} {115}},\ \bibinfo {pages} {098302} (\bibinfo {year} {2015})}\BibitemShut {NoStop}%
\bibitem [{\citenamefont {Utter}\ and\ \citenamefont {Behringer}(2008)}]{Utter2008AffineNonaffineGranularShear}%
  \BibitemOpen
  \bibfield  {author} {\bibinfo {author} {\bibfnamefont {B.}~\bibnamefont {Utter}}\ and\ \bibinfo {author} {\bibfnamefont {R.~P.}\ \bibnamefont {Behringer}},\ }\bibfield  {title} {\bibinfo {title} {Experimental measures of affine and nonaffine deformation in granular shear},\ }\href {https://doi.org/10.1103/PhysRevLett.100.208302} {\bibfield  {journal} {\bibinfo  {journal} {Physical Review Letters}\ }\textbf {\bibinfo {volume} {100}},\ \bibinfo {pages} {208302} (\bibinfo {year} {2008})}\BibitemShut {NoStop}%
\bibitem [{\citenamefont {Chen}\ \emph {et~al.}(2010)\citenamefont {Chen}, \citenamefont {Semwogerere}, \citenamefont {Sato}, \citenamefont {Breedveld},\ and\ \citenamefont {Weeks}}]{Chen2010ShearedColloidalLiquid}%
  \BibitemOpen
  \bibfield  {author} {\bibinfo {author} {\bibfnamefont {D.}~\bibnamefont {Chen}}, \bibinfo {author} {\bibfnamefont {D.}~\bibnamefont {Semwogerere}}, \bibinfo {author} {\bibfnamefont {J.}~\bibnamefont {Sato}}, \bibinfo {author} {\bibfnamefont {V.}~\bibnamefont {Breedveld}},\ and\ \bibinfo {author} {\bibfnamefont {E.~R.}\ \bibnamefont {Weeks}},\ }\bibfield  {title} {\bibinfo {title} {Microscopic structural relaxation in a sheared supercooled colloidal liquid},\ }\href {https://doi.org/10.1103/PhysRevE.81.011403} {\bibfield  {journal} {\bibinfo  {journal} {Physical Review E}\ }\textbf {\bibinfo {volume} {81}},\ \bibinfo {pages} {011403} (\bibinfo {year} {2010})}\BibitemShut {NoStop}%
\bibitem [{\citenamefont {Besseling}\ \emph {et~al.}(2007)\citenamefont {Besseling}, \citenamefont {Weeks}, \citenamefont {Schofield},\ and\ \citenamefont {Poon}}]{besseling07}%
  \BibitemOpen
  \bibfield  {author} {\bibinfo {author} {\bibfnamefont {R.}~\bibnamefont {Besseling}}, \bibinfo {author} {\bibfnamefont {E.~R.}\ \bibnamefont {Weeks}}, \bibinfo {author} {\bibfnamefont {A.~B.}\ \bibnamefont {Schofield}},\ and\ \bibinfo {author} {\bibfnamefont {W.~C.~K.}\ \bibnamefont {Poon}},\ }\bibfield  {title} {\bibinfo {title} {{Three-Dimensional} imaging of colloidal glasses under steady shear},\ }\href {https://doi.org/10.1103/physrevlett.99.028301} {\bibfield  {journal} {\bibinfo  {journal} {Phys. Rev. Lett.}\ }\textbf {\bibinfo {volume} {99}},\ \bibinfo {pages} {028301} (\bibinfo {year} {2007})}\BibitemShut {NoStop}%
\bibitem [{\citenamefont {Falk}\ and\ \citenamefont {Langer}(1998)}]{FalkLanger1998}%
  \BibitemOpen
  \bibfield  {author} {\bibinfo {author} {\bibfnamefont {M.~L.}\ \bibnamefont {Falk}}\ and\ \bibinfo {author} {\bibfnamefont {J.~S.}\ \bibnamefont {Langer}},\ }\bibfield  {title} {\bibinfo {title} {Dynamics of viscoplastic deformation in amorphous solids},\ }\href {https://doi.org/10.1103/PhysRevE.57.7192} {\bibfield  {journal} {\bibinfo  {journal} {Phys. Rev. E}\ }\textbf {\bibinfo {volume} {57}},\ \bibinfo {pages} {7192} (\bibinfo {year} {1998})}\BibitemShut {NoStop}%
\bibitem [{\citenamefont {Liu}\ \emph {et~al.}(1996)\citenamefont {Liu}, \citenamefont {Ramaswamy}, \citenamefont {Mason}, \citenamefont {Gang},\ and\ \citenamefont {Weitz}}]{liu96}%
  \BibitemOpen
  \bibfield  {author} {\bibinfo {author} {\bibfnamefont {A.~J.}\ \bibnamefont {Liu}}, \bibinfo {author} {\bibfnamefont {S.}~\bibnamefont {Ramaswamy}}, \bibinfo {author} {\bibfnamefont {T.~G.}\ \bibnamefont {Mason}}, \bibinfo {author} {\bibfnamefont {H.}~\bibnamefont {Gang}},\ and\ \bibinfo {author} {\bibfnamefont {D.~A.}\ \bibnamefont {Weitz}},\ }\bibfield  {title} {\bibinfo {title} {Anomalous viscous loss in emulsions},\ }\href {https://doi.org/10.1103/physrevlett.76.3017} {\bibfield  {journal} {\bibinfo  {journal} {Phys. Rev. Lett.}\ }\textbf {\bibinfo {volume} {76}},\ \bibinfo {pages} {3017} (\bibinfo {year} {1996})}\BibitemShut {NoStop}%
\bibitem [{\citenamefont {Mason}\ \emph {et~al.}(1997)\citenamefont {Mason}, \citenamefont {Lacasse}, \citenamefont {Grest}, \citenamefont {Levine}, \citenamefont {Bibette},\ and\ \citenamefont {Weitz}}]{mason97emulsions}%
  \BibitemOpen
  \bibfield  {author} {\bibinfo {author} {\bibfnamefont {T.~G.}\ \bibnamefont {Mason}}, \bibinfo {author} {\bibfnamefont {M.-D.}\ \bibnamefont {Lacasse}}, \bibinfo {author} {\bibfnamefont {G.~S.}\ \bibnamefont {Grest}}, \bibinfo {author} {\bibfnamefont {D.}~\bibnamefont {Levine}}, \bibinfo {author} {\bibfnamefont {J.}~\bibnamefont {Bibette}},\ and\ \bibinfo {author} {\bibfnamefont {D.~A.}\ \bibnamefont {Weitz}},\ }\bibfield  {title} {\bibinfo {title} {Osmotic pressure and viscoelastic shear moduli of concentrated emulsions},\ }\href {https://doi.org/10.1103/physreve.56.3150} {\bibfield  {journal} {\bibinfo  {journal} {Phys. Rev. E}\ }\textbf {\bibinfo {volume} {56}},\ \bibinfo {pages} {3150} (\bibinfo {year} {1997})}\BibitemShut {NoStop}%
\bibitem [{\citenamefont {H\'{e}braud}\ \emph {et~al.}(1997)\citenamefont {H\'{e}braud}, \citenamefont {Lequeux}, \citenamefont {Munch},\ and\ \citenamefont {Pine}}]{hebraud97}%
  \BibitemOpen
  \bibfield  {author} {\bibinfo {author} {\bibfnamefont {P.}~\bibnamefont {H\'{e}braud}}, \bibinfo {author} {\bibfnamefont {F.}~\bibnamefont {Lequeux}}, \bibinfo {author} {\bibfnamefont {J.~P.}\ \bibnamefont {Munch}},\ and\ \bibinfo {author} {\bibfnamefont {D.~J.}\ \bibnamefont {Pine}},\ }\bibfield  {title} {\bibinfo {title} {Yielding and rearrangements in disordered emulsions},\ }\href {https://doi.org/10.1103/physrevlett.78.4657} {\bibfield  {journal} {\bibinfo  {journal} {Phys. Rev. Lett.}\ }\textbf {\bibinfo {volume} {78}},\ \bibinfo {pages} {4657} (\bibinfo {year} {1997})}\BibitemShut {NoStop}%
\bibitem [{\citenamefont {Petekidis}\ \emph {et~al.}(2002)\citenamefont {Petekidis}, \citenamefont {Moussa{\"{\i}}d},\ and\ \citenamefont {Pusey}}]{petekidis02}%
  \BibitemOpen
  \bibfield  {author} {\bibinfo {author} {\bibfnamefont {G.}~\bibnamefont {Petekidis}}, \bibinfo {author} {\bibfnamefont {A.}~\bibnamefont {Moussa{\"{\i}}d}},\ and\ \bibinfo {author} {\bibfnamefont {P.~N.}\ \bibnamefont {Pusey}},\ }\bibfield  {title} {\bibinfo {title} {Rearrangements in hard-sphere glasses under oscillatory shear strain},\ }\href {https://doi.org/10.1103/physreve.66.051402} {\bibfield  {journal} {\bibinfo  {journal} {Phys. Rev. E}\ }\textbf {\bibinfo {volume} {66}},\ \bibinfo {pages} {051402} (\bibinfo {year} {2002})}\BibitemShut {NoStop}%
\bibitem [{\citenamefont {Schall}\ \emph {et~al.}(2007)\citenamefont {Schall}, \citenamefont {Weitz},\ and\ \citenamefont {Spaepen}}]{schall07}%
  \BibitemOpen
  \bibfield  {author} {\bibinfo {author} {\bibfnamefont {P.}~\bibnamefont {Schall}}, \bibinfo {author} {\bibfnamefont {D.~A.}\ \bibnamefont {Weitz}},\ and\ \bibinfo {author} {\bibfnamefont {F.}~\bibnamefont {Spaepen}},\ }\bibfield  {title} {\bibinfo {title} {Structural rearrangements that govern flow in colloidal glasses},\ }\href {https://doi.org/10.1126/science.1149308} {\bibfield  {journal} {\bibinfo  {journal} {Science}\ }\textbf {\bibinfo {volume} {318}},\ \bibinfo {pages} {1895} (\bibinfo {year} {2007})}\BibitemShut {NoStop}%
\bibitem [{\citenamefont {Vasisht}\ \emph {et~al.}(2018)\citenamefont {Vasisht}, \citenamefont {Dutta}, \citenamefont {Del~Gado},\ and\ \citenamefont {Blair}}]{vasisht18}%
  \BibitemOpen
  \bibfield  {author} {\bibinfo {author} {\bibfnamefont {V.~V.}\ \bibnamefont {Vasisht}}, \bibinfo {author} {\bibfnamefont {S.~K.}\ \bibnamefont {Dutta}}, \bibinfo {author} {\bibfnamefont {E.}~\bibnamefont {Del~Gado}},\ and\ \bibinfo {author} {\bibfnamefont {D.~L.}\ \bibnamefont {Blair}},\ }\bibfield  {title} {\bibinfo {title} {Rate dependence of elementary rearrangements and spatiotemporal correlations in the {3D} flow of soft solids},\ }\bibfield  {journal} {\bibinfo  {journal} {Phys. Rev. Lett.}\ }\textbf {\bibinfo {volume} {120}},\ \href {https://doi.org/10.1103/physrevlett.120.018001} {\bibinfo {pages} {018001}} (\bibinfo {year} {2018})\BibitemShut {NoStop}%
\bibitem [{\citenamefont {Tsai}\ \emph {et~al.}(2021)\citenamefont {Tsai}, \citenamefont {Huang},\ and\ \citenamefont {Tsai}}]{tsai21}%
  \BibitemOpen
  \bibfield  {author} {\bibinfo {author} {\bibfnamefont {J.-C.~J.}\ \bibnamefont {Tsai}}, \bibinfo {author} {\bibfnamefont {G.-H.}\ \bibnamefont {Huang}},\ and\ \bibinfo {author} {\bibfnamefont {C.-E.}\ \bibnamefont {Tsai}},\ }\bibfield  {title} {\bibinfo {title} {Signature of transition between granular solid and fluid: {Rate}-dependent stick slips in steady shearing},\ }\href {https://doi.org/10.1103/PhysRevLett.126.128001} {\bibfield  {journal} {\bibinfo  {journal} {Phys. Rev. Lett.}\ }\textbf {\bibinfo {volume} {126}},\ \bibinfo {pages} {128001} (\bibinfo {year} {2021})}\BibitemShut {NoStop}%
\bibitem [{\citenamefont {Yamamoto}\ and\ \citenamefont {Onuki}(1997)}]{yamamoto97}%
  \BibitemOpen
  \bibfield  {author} {\bibinfo {author} {\bibfnamefont {R.}~\bibnamefont {Yamamoto}}\ and\ \bibinfo {author} {\bibfnamefont {A.}~\bibnamefont {Onuki}},\ }\bibfield  {title} {\bibinfo {title} {Nonlinear rheology of a highly supercooled liquid},\ }\href@noop {} {\bibfield  {journal} {\bibinfo  {journal} {Europhys. Lett.}\ }\textbf {\bibinfo {volume} {40}},\ \bibinfo {pages} {61} (\bibinfo {year} {1997})}\BibitemShut {NoStop}%
\bibitem [{\citenamefont {Olsson}\ and\ \citenamefont {Teitel}(2007)}]{teitel07}%
  \BibitemOpen
  \bibfield  {author} {\bibinfo {author} {\bibfnamefont {P.}~\bibnamefont {Olsson}}\ and\ \bibinfo {author} {\bibfnamefont {S.}~\bibnamefont {Teitel}},\ }\bibfield  {title} {\bibinfo {title} {Critical scaling of shear viscosity at the jamming transition},\ }\href {https://doi.org/10.1103/physrevlett.99.178001} {\bibfield  {journal} {\bibinfo  {journal} {Phys. Rev. Lett.}\ }\textbf {\bibinfo {volume} {99}},\ \bibinfo {pages} {178001} (\bibinfo {year} {2007})}\BibitemShut {NoStop}%
\bibitem [{\citenamefont {Lema{\^\i}tre}\ and\ \citenamefont {Caroli}(2009)}]{lemaitre09}%
  \BibitemOpen
  \bibfield  {author} {\bibinfo {author} {\bibfnamefont {A.}~\bibnamefont {Lema{\^\i}tre}}\ and\ \bibinfo {author} {\bibfnamefont {C.}~\bibnamefont {Caroli}},\ }\bibfield  {title} {\bibinfo {title} {{Rate-Dependent} avalanche size in athermally sheared amorphous solids},\ }\href {https://doi.org/10.1103/physrevlett.103.065501} {\bibfield  {journal} {\bibinfo  {journal} {Phys. Rev. Lett.}\ }\textbf {\bibinfo {volume} {103}},\ \bibinfo {pages} {065501} (\bibinfo {year} {2009})}\BibitemShut {NoStop}%
\bibitem [{\citenamefont {Manning}\ and\ \citenamefont {Liu}(2011)}]{manning11}%
  \BibitemOpen
  \bibfield  {author} {\bibinfo {author} {\bibfnamefont {M.~L.}\ \bibnamefont {Manning}}\ and\ \bibinfo {author} {\bibfnamefont {A.~J.}\ \bibnamefont {Liu}},\ }\bibfield  {title} {\bibinfo {title} {Vibrational modes identify soft spots in a sheared disordered packing},\ }\href {https://doi.org/10.1103/physrevlett.107.108302} {\bibfield  {journal} {\bibinfo  {journal} {Phys. Rev. Lett.}\ }\textbf {\bibinfo {volume} {107}},\ \bibinfo {pages} {108302} (\bibinfo {year} {2011})}\BibitemShut {NoStop}%
\bibitem [{\citenamefont {Cubuk}\ \emph {et~al.}(2015)\citenamefont {Cubuk}, \citenamefont {Schoenholz}, \citenamefont {Rieser}, \citenamefont {Malone}, \citenamefont {Rottler}, \citenamefont {Durian}, \citenamefont {Kaxiras},\ and\ \citenamefont {Liu}}]{cubuk15}%
  \BibitemOpen
  \bibfield  {author} {\bibinfo {author} {\bibfnamefont {E.~D.}\ \bibnamefont {Cubuk}}, \bibinfo {author} {\bibfnamefont {S.~S.}\ \bibnamefont {Schoenholz}}, \bibinfo {author} {\bibfnamefont {J.~M.}\ \bibnamefont {Rieser}}, \bibinfo {author} {\bibfnamefont {B.~D.}\ \bibnamefont {Malone}}, \bibinfo {author} {\bibfnamefont {J.}~\bibnamefont {Rottler}}, \bibinfo {author} {\bibfnamefont {D.~J.}\ \bibnamefont {Durian}}, \bibinfo {author} {\bibfnamefont {E.}~\bibnamefont {Kaxiras}},\ and\ \bibinfo {author} {\bibfnamefont {A.~J.}\ \bibnamefont {Liu}},\ }\bibfield  {title} {\bibinfo {title} {Identifying structural flow defects in disordered solids using machine-learning methods},\ }\href {https://doi.org/10.1103/physrevlett.114.108001} {\bibfield  {journal} {\bibinfo  {journal} {Phys. Rev. Lett.}\ }\textbf {\bibinfo {volume} {114}},\ \bibinfo {pages} {108001} (\bibinfo {year} {2015})}\BibitemShut {NoStop}%
\bibitem [{\citenamefont {Hassani}\ \emph {et~al.}(2019)\citenamefont {Hassani}, \citenamefont {Lagogianni},\ and\ \citenamefont {Varnik}}]{hassani19}%
  \BibitemOpen
  \bibfield  {author} {\bibinfo {author} {\bibfnamefont {M.}~\bibnamefont {Hassani}}, \bibinfo {author} {\bibfnamefont {A.~E.}\ \bibnamefont {Lagogianni}},\ and\ \bibinfo {author} {\bibfnamefont {F.}~\bibnamefont {Varnik}},\ }\bibfield  {title} {\bibinfo {title} {Probing the degree of heterogeneity within a shear band of a model glass},\ }\href {https://doi.org/10.1103/PhysRevLett.123.195502} {\bibfield  {journal} {\bibinfo  {journal} {Phys. Rev. Lett.}\ }\textbf {\bibinfo {volume} {123}},\ \bibinfo {pages} {195502} (\bibinfo {year} {2019})}\BibitemShut {NoStop}%
\bibitem [{\citenamefont {Losert}\ \emph {et~al.}(2000)\citenamefont {Losert}, \citenamefont {Bocquet}, \citenamefont {Lubensky},\ and\ \citenamefont {Gollub}}]{losert00}%
  \BibitemOpen
  \bibfield  {author} {\bibinfo {author} {\bibfnamefont {W.}~\bibnamefont {Losert}}, \bibinfo {author} {\bibfnamefont {L.}~\bibnamefont {Bocquet}}, \bibinfo {author} {\bibfnamefont {T.~C.}\ \bibnamefont {Lubensky}},\ and\ \bibinfo {author} {\bibfnamefont {J.~P.}\ \bibnamefont {Gollub}},\ }\bibfield  {title} {\bibinfo {title} {Particle dynamics in sheared granular matter},\ }\href {https://doi.org/10.1103/physrevlett.85.1428} {\bibfield  {journal} {\bibinfo  {journal} {Phys. Rev. Lett.}\ }\textbf {\bibinfo {volume} {85}},\ \bibinfo {pages} {1428} (\bibinfo {year} {2000})}\BibitemShut {NoStop}%
\bibitem [{\citenamefont {Patinet}\ \emph {et~al.}(2016)\citenamefont {Patinet}, \citenamefont {Vandembroucq},\ and\ \citenamefont {Falk}}]{patinet2016connecting}%
  \BibitemOpen
  \bibfield  {author} {\bibinfo {author} {\bibfnamefont {S.}~\bibnamefont {Patinet}}, \bibinfo {author} {\bibfnamefont {D.}~\bibnamefont {Vandembroucq}},\ and\ \bibinfo {author} {\bibfnamefont {M.~L.}\ \bibnamefont {Falk}},\ }\bibfield  {title} {\bibinfo {title} {Connecting local yield stresses with plastic activity in amorphous solids},\ }\href@noop {} {\bibfield  {journal} {\bibinfo  {journal} {Phys. Rev. Lett.}\ }\textbf {\bibinfo {volume} {117}},\ \bibinfo {pages} {045501} (\bibinfo {year} {2016})}\BibitemShut {NoStop}%
\bibitem [{\citenamefont {Chen}\ \emph {et~al.}(2015)\citenamefont {Chen}, \citenamefont {Desmond},\ and\ \citenamefont {Weeks}}]{chen15}%
  \BibitemOpen
  \bibfield  {author} {\bibinfo {author} {\bibfnamefont {D.}~\bibnamefont {Chen}}, \bibinfo {author} {\bibfnamefont {K.~W.}\ \bibnamefont {Desmond}},\ and\ \bibinfo {author} {\bibfnamefont {E.~R.}\ \bibnamefont {Weeks}},\ }\bibfield  {title} {\bibinfo {title} {Experimental observation of local rearrangements in dense quasi-two-dimensional emulsion flow},\ }\href {https://doi.org/10.1103/physreve.91.062306} {\bibfield  {journal} {\bibinfo  {journal} {Phys. Rev. E}\ }\textbf {\bibinfo {volume} {91}},\ \bibinfo {pages} {062306} (\bibinfo {year} {2015})}\BibitemShut {NoStop}%
\bibitem [{\citenamefont {Jiang}\ \emph {et~al.}(2019)\citenamefont {Jiang}, \citenamefont {Weeks},\ and\ \citenamefont {Bailey}}]{jiang19}%
  \BibitemOpen
  \bibfield  {author} {\bibinfo {author} {\bibfnamefont {Y.}~\bibnamefont {Jiang}}, \bibinfo {author} {\bibfnamefont {E.~R.}\ \bibnamefont {Weeks}},\ and\ \bibinfo {author} {\bibfnamefont {N.~P.}\ \bibnamefont {Bailey}},\ }\bibfield  {title} {\bibinfo {title} {Isomorph invariance of dynamics of sheared glassy systems},\ }\href {https://doi.org/10.1103/PhysRevE.100.053005} {\bibfield  {journal} {\bibinfo  {journal} {Phys. Rev. E}\ }\textbf {\bibinfo {volume} {100}},\ \bibinfo {pages} {053005} (\bibinfo {year} {2019})}\BibitemShut {NoStop}%
\bibitem [{\citenamefont {Haeberli}\ \emph {et~al.}(2006)\citenamefont {Haeberli}, \citenamefont {Hallet}, \citenamefont {Arenson}, \citenamefont {Elconin}, \citenamefont {Humlum}, \citenamefont {K\"{a}\"{a}b}, \citenamefont {Kaufmann}, \citenamefont {Ladanyi}, \citenamefont {Matsuoka}, \citenamefont {Springman},\ and\ \citenamefont {M\"{u}hll}}]{haeberli06}%
  \BibitemOpen
  \bibfield  {author} {\bibinfo {author} {\bibfnamefont {W.}~\bibnamefont {Haeberli}}, \bibinfo {author} {\bibfnamefont {B.}~\bibnamefont {Hallet}}, \bibinfo {author} {\bibfnamefont {L.}~\bibnamefont {Arenson}}, \bibinfo {author} {\bibfnamefont {R.}~\bibnamefont {Elconin}}, \bibinfo {author} {\bibfnamefont {O.~e.}\ \bibnamefont {Humlum}}, \bibinfo {author} {\bibfnamefont {A.}~\bibnamefont {K\"{a}\"{a}b}}, \bibinfo {author} {\bibfnamefont {V.}~\bibnamefont {Kaufmann}}, \bibinfo {author} {\bibfnamefont {B.}~\bibnamefont {Ladanyi}}, \bibinfo {author} {\bibfnamefont {N.}~\bibnamefont {Matsuoka}}, \bibinfo {author} {\bibfnamefont {S.}~\bibnamefont {Springman}},\ and\ \bibinfo {author} {\bibfnamefont {D.~V.}\ \bibnamefont {M\"{u}hll}},\ }\bibfield  {title} {\bibinfo {title} {Permafrost creep and rock glacier dynamics},\ }\href {https://doi.org/10.1002/ppp.561} {\bibfield  {journal} {\bibinfo  {journal} {Permafrost Periglac. Process.}\ }\textbf {\bibinfo {volume} {17}},\ \bibinfo {pages}
  {189} (\bibinfo {year} {2006})}\BibitemShut {NoStop}%
\bibitem [{\citenamefont {Pitman}\ and\ \citenamefont {Le}(2005)}]{pitman05}%
  \BibitemOpen
  \bibfield  {author} {\bibinfo {author} {\bibfnamefont {E.~B.}\ \bibnamefont {Pitman}}\ and\ \bibinfo {author} {\bibfnamefont {L.}~\bibnamefont {Le}},\ }\bibfield  {title} {\bibinfo {title} {A two-fluid model for avalanche and debris flows},\ }\href {https://doi.org/10.1098/rsta.2005.1596} {\bibfield  {journal} {\bibinfo  {journal} {Phil. Trans. Royal Soc. London A}\ }\textbf {\bibinfo {volume} {363}},\ \bibinfo {pages} {1573} (\bibinfo {year} {2005})}\BibitemShut {NoStop}%
\bibitem [{\citenamefont {Burton}\ \emph {et~al.}(2018)\citenamefont {Burton}, \citenamefont {Amundson}, \citenamefont {Cassotto}, \citenamefont {Kuo},\ and\ \citenamefont {Dennin}}]{burton18}%
  \BibitemOpen
  \bibfield  {author} {\bibinfo {author} {\bibfnamefont {J.~C.}\ \bibnamefont {Burton}}, \bibinfo {author} {\bibfnamefont {J.~M.}\ \bibnamefont {Amundson}}, \bibinfo {author} {\bibfnamefont {R.}~\bibnamefont {Cassotto}}, \bibinfo {author} {\bibfnamefont {C.-C.}\ \bibnamefont {Kuo}},\ and\ \bibinfo {author} {\bibfnamefont {M.}~\bibnamefont {Dennin}},\ }\bibfield  {title} {\bibinfo {title} {Quantifying flow and stress in ice mélange, the world’s largest granular material},\ }\href {https://doi.org/10.1073/pnas.1715136115} {\bibfield  {journal} {\bibinfo  {journal} {Proc. Nat. Acad. Sci.}\ }\textbf {\bibinfo {volume} {115}},\ \bibinfo {pages} {5105} (\bibinfo {year} {2018})}\BibitemShut {NoStop}%
\bibitem [{\citenamefont {Or}\ and\ \citenamefont {Ghezzehei}(2002)}]{or02}%
  \BibitemOpen
  \bibfield  {author} {\bibinfo {author} {\bibfnamefont {D.}~\bibnamefont {Or}}\ and\ \bibinfo {author} {\bibfnamefont {T.~A.}\ \bibnamefont {Ghezzehei}},\ }\bibfield  {title} {\bibinfo {title} {Modeling post-tillage soil structural dynamics: a review},\ }\href {https://doi.org/10.1016/s0167-1987(01)00256-2} {\bibfield  {journal} {\bibinfo  {journal} {Soil and Tillage Research}\ }\textbf {\bibinfo {volume} {64}},\ \bibinfo {pages} {41} (\bibinfo {year} {2002})}\BibitemShut {NoStop}%
\bibitem [{\citenamefont {Besq}\ \emph {et~al.}(2003)\citenamefont {Besq}, \citenamefont {Malfoy}, \citenamefont {Pantet}, \citenamefont {Monnet},\ and\ \citenamefont {Righi}}]{besq03}%
  \BibitemOpen
  \bibfield  {author} {\bibinfo {author} {\bibfnamefont {A.}~\bibnamefont {Besq}}, \bibinfo {author} {\bibfnamefont {C.}~\bibnamefont {Malfoy}}, \bibinfo {author} {\bibfnamefont {A.}~\bibnamefont {Pantet}}, \bibinfo {author} {\bibfnamefont {P.}~\bibnamefont {Monnet}},\ and\ \bibinfo {author} {\bibfnamefont {D.}~\bibnamefont {Righi}},\ }\bibfield  {title} {\bibinfo {title} {Physicochemical characterisation and flow properties of some bentonite muds},\ }\href {https://doi.org/10.1016/s0169-1317(03)00127-3} {\bibfield  {journal} {\bibinfo  {journal} {Applied Clay Science}\ }\textbf {\bibinfo {volume} {23}},\ \bibinfo {pages} {275} (\bibinfo {year} {2003})}\BibitemShut {NoStop}%
\bibitem [{\citenamefont {Rosquo\"{e}t}\ \emph {et~al.}(2003)\citenamefont {Rosquo\"{e}t}, \citenamefont {Alexis}, \citenamefont {Khelidj},\ and\ \citenamefont {Phelipot}}]{rosquoet03}%
  \BibitemOpen
  \bibfield  {author} {\bibinfo {author} {\bibfnamefont {F.}~\bibnamefont {Rosquo\"{e}t}}, \bibinfo {author} {\bibfnamefont {A.}~\bibnamefont {Alexis}}, \bibinfo {author} {\bibfnamefont {A.}~\bibnamefont {Khelidj}},\ and\ \bibinfo {author} {\bibfnamefont {A.}~\bibnamefont {Phelipot}},\ }\bibfield  {title} {\bibinfo {title} {Experimental study of cement grout},\ }\href {https://doi.org/10.1016/s0008-8846(02)01036-0} {\bibfield  {journal} {\bibinfo  {journal} {Cement and Concrete Research}\ }\textbf {\bibinfo {volume} {33}},\ \bibinfo {pages} {713} (\bibinfo {year} {2003})}\BibitemShut {NoStop}%
\bibitem [{\citenamefont {Taylor}\ \emph {et~al.}(2009)\citenamefont {Taylor}, \citenamefont {Van~Damme}, \citenamefont {Johns}, \citenamefont {Routh},\ and\ \citenamefont {Wilson}}]{taylor09}%
  \BibitemOpen
  \bibfield  {author} {\bibinfo {author} {\bibfnamefont {J.~E.}\ \bibnamefont {Taylor}}, \bibinfo {author} {\bibfnamefont {I.}~\bibnamefont {Van~Damme}}, \bibinfo {author} {\bibfnamefont {M.~L.}\ \bibnamefont {Johns}}, \bibinfo {author} {\bibfnamefont {A.~F.}\ \bibnamefont {Routh}},\ and\ \bibinfo {author} {\bibfnamefont {D.~I.}\ \bibnamefont {Wilson}},\ }\bibfield  {title} {\bibinfo {title} {Shear rheology of molten crumb chocolate},\ }\href {https://doi.org/10.1111/j.1750-3841.2008.01041.x} {\bibfield  {journal} {\bibinfo  {journal} {J. Food Sci.}\ }\textbf {\bibinfo {volume} {74}},\ \bibinfo {pages} {E55} (\bibinfo {year} {2009})}\BibitemShut {NoStop}%
\bibitem [{\citenamefont {Lynch}\ \emph {et~al.}(2008)\citenamefont {Lynch}, \citenamefont {Cianci},\ and\ \citenamefont {Weeks}}]{Lynch2008AgingBinaryColloidalGlass}%
  \BibitemOpen
  \bibfield  {author} {\bibinfo {author} {\bibfnamefont {J.~M.}\ \bibnamefont {Lynch}}, \bibinfo {author} {\bibfnamefont {G.~C.}\ \bibnamefont {Cianci}},\ and\ \bibinfo {author} {\bibfnamefont {E.~R.}\ \bibnamefont {Weeks}},\ }\bibfield  {title} {\bibinfo {title} {Dynamics and structure of an aging binary colloidal glass},\ }\href {https://doi.org/10.1103/PhysRevE.78.031410} {\bibfield  {journal} {\bibinfo  {journal} {Physical Review E}\ }\textbf {\bibinfo {volume} {78}},\ \bibinfo {pages} {031410} (\bibinfo {year} {2008})}\BibitemShut {NoStop}%
\bibitem [{\citenamefont {Jiang}\ \emph {et~al.}(2023)\citenamefont {Jiang}, \citenamefont {Sussman},\ and\ \citenamefont {Weeks}}]{JiangSussmanWeeks2023}%
  \BibitemOpen
  \bibfield  {author} {\bibinfo {author} {\bibfnamefont {Y.}~\bibnamefont {Jiang}}, \bibinfo {author} {\bibfnamefont {D.~M.}\ \bibnamefont {Sussman}},\ and\ \bibinfo {author} {\bibfnamefont {E.~R.}\ \bibnamefont {Weeks}},\ }\bibfield  {title} {\bibinfo {title} {Effects of polydispersity on the plastic behaviors of dense two-dimensional granular systems under shear},\ }\href {https://doi.org/10.1103/PhysRevE.108.054605} {\bibfield  {journal} {\bibinfo  {journal} {Phys. Rev. E}\ }\textbf {\bibinfo {volume} {108}},\ \bibinfo {pages} {054605} (\bibinfo {year} {2023})}\BibitemShut {NoStop}%
\bibitem [{\citenamefont {Illing}\ and\ \citenamefont {Weeks}(2025)}]{IllingWeeks2025NonaffinePolydisperse}%
  \BibitemOpen
  \bibfield  {author} {\bibinfo {author} {\bibfnamefont {P.~E.}\ \bibnamefont {Illing}}\ and\ \bibinfo {author} {\bibfnamefont {E.~R.}\ \bibnamefont {Weeks}},\ }\bibfield  {title} {\bibinfo {title} {Nonaffine motion in flowing highly polydisperse granular media},\ }\href {https://doi.org/10.1103/PhysRevE.111.045422} {\bibfield  {journal} {\bibinfo  {journal} {Phys. Rev. E}\ }\textbf {\bibinfo {volume} {111}},\ \bibinfo {pages} {045422} (\bibinfo {year} {2025})}\BibitemShut {NoStop}%
\bibitem [{\citenamefont {van Meel}\ \emph {et~al.}(2012)\citenamefont {van Meel}, \citenamefont {Filion}, \citenamefont {Valeriani},\ and\ \citenamefont {Frenkel}}]{vanMeel2012}%
  \BibitemOpen
  \bibfield  {author} {\bibinfo {author} {\bibfnamefont {J.~A.}\ \bibnamefont {van Meel}}, \bibinfo {author} {\bibfnamefont {L.}~\bibnamefont {Filion}}, \bibinfo {author} {\bibfnamefont {C.}~\bibnamefont {Valeriani}},\ and\ \bibinfo {author} {\bibfnamefont {D.}~\bibnamefont {Frenkel}},\ }\bibfield  {title} {\bibinfo {title} {A parameter-free, solid-angle based, nearest-neighbor algorithm},\ }\href@noop {} {\bibfield  {journal} {\bibinfo  {journal} {The Journal of Chemical Physics}\ }\textbf {\bibinfo {volume} {136}},\ \bibinfo {pages} {234107} (\bibinfo {year} {2012})}\BibitemShut {NoStop}%
\bibitem [{\citenamefont {Durian}(1995)}]{durian95}%
  \BibitemOpen
  \bibfield  {author} {\bibinfo {author} {\bibfnamefont {D.~J.}\ \bibnamefont {Durian}},\ }\bibfield  {title} {\bibinfo {title} {Foam mechanics at the bubble scale},\ }\href {https://doi.org/10.1103/physrevlett.75.4780} {\bibfield  {journal} {\bibinfo  {journal} {Phys. Rev. Lett.}\ }\textbf {\bibinfo {volume} {75}},\ \bibinfo {pages} {4780} (\bibinfo {year} {1995})}\BibitemShut {NoStop}%
\bibitem [{\citenamefont {Tewari}\ \emph {et~al.}(1999)\citenamefont {Tewari}, \citenamefont {Schiemann}, \citenamefont {Durian}, \citenamefont {Knobler}, \citenamefont {Langer},\ and\ \citenamefont {Liu}}]{tewari99}%
  \BibitemOpen
  \bibfield  {author} {\bibinfo {author} {\bibfnamefont {S.}~\bibnamefont {Tewari}}, \bibinfo {author} {\bibfnamefont {D.}~\bibnamefont {Schiemann}}, \bibinfo {author} {\bibfnamefont {D.~J.}\ \bibnamefont {Durian}}, \bibinfo {author} {\bibfnamefont {C.~M.}\ \bibnamefont {Knobler}}, \bibinfo {author} {\bibfnamefont {S.~A.}\ \bibnamefont {Langer}},\ and\ \bibinfo {author} {\bibfnamefont {A.~J.}\ \bibnamefont {Liu}},\ }\bibfield  {title} {\bibinfo {title} {Statistics of shear-induced rearrangements in a two-dimensional model foam},\ }\href {https://doi.org/10.1103/physreve.60.4385} {\bibfield  {journal} {\bibinfo  {journal} {Phys. Rev. E}\ }\textbf {\bibinfo {volume} {60}},\ \bibinfo {pages} {4385} (\bibinfo {year} {1999})}\BibitemShut {NoStop}%
\bibitem [{\citenamefont {Chikkadi}\ and\ \citenamefont {Schall}(2012)}]{ChikkadiSchall2012}%
  \BibitemOpen
  \bibfield  {author} {\bibinfo {author} {\bibfnamefont {V.}~\bibnamefont {Chikkadi}}\ and\ \bibinfo {author} {\bibfnamefont {P.}~\bibnamefont {Schall}},\ }\bibfield  {title} {\bibinfo {title} {Nonaffine measures of particle displacements in sheared colloidal glasses},\ }\href {https://doi.org/10.1103/PhysRevE.85.031402} {\bibfield  {journal} {\bibinfo  {journal} {Phys. Rev. E}\ }\textbf {\bibinfo {volume} {85}},\ \bibinfo {pages} {031402} (\bibinfo {year} {2012})}\BibitemShut {NoStop}%
\bibitem [{\citenamefont {Preparata}\ and\ \citenamefont {Shamos}(1985)}]{preparata85}%
  \BibitemOpen
  \bibfield  {author} {\bibinfo {author} {\bibfnamefont {F.~P.}\ \bibnamefont {Preparata}}\ and\ \bibinfo {author} {\bibfnamefont {M.~I.}\ \bibnamefont {Shamos}},\ }\href@noop {} {\emph {\bibinfo {title} {Computational Geometry}}}\ (\bibinfo  {publisher} {Springer-Verlag, New York},\ \bibinfo {year} {1985})\BibitemShut {NoStop}%
\bibitem [{\citenamefont {Okabe}\ \emph {et~al.}(2000)\citenamefont {Okabe}, \citenamefont {Boots}, \citenamefont {Sugihara},\ and\ \citenamefont {Chiu}}]{Okabe2000}%
  \BibitemOpen
  \bibfield  {author} {\bibinfo {author} {\bibfnamefont {A.}~\bibnamefont {Okabe}}, \bibinfo {author} {\bibfnamefont {B.}~\bibnamefont {Boots}}, \bibinfo {author} {\bibfnamefont {K.}~\bibnamefont {Sugihara}},\ and\ \bibinfo {author} {\bibfnamefont {S.~N.}\ \bibnamefont {Chiu}},\ }\href@noop {} {\emph {\bibinfo {title} {Spatial Tessellations: Concepts and Applications of Voronoi Diagrams}}}\ (\bibinfo  {publisher} {Wiley},\ \bibinfo {year} {2000})\BibitemShut {NoStop}%
\bibitem [{\citenamefont {Aurenhammer}(1987)}]{Aurenhammer1987}%
  \BibitemOpen
  \bibfield  {author} {\bibinfo {author} {\bibfnamefont {F.}~\bibnamefont {Aurenhammer}},\ }\bibfield  {title} {\bibinfo {title} {Power diagrams: Properties, algorithms and applications},\ }\href {https://doi.org/10.1137/0216006} {\bibfield  {journal} {\bibinfo  {journal} {SIAM Journal on Computing}\ }\textbf {\bibinfo {volume} {16}},\ \bibinfo {pages} {78} (\bibinfo {year} {1987})}\BibitemShut {NoStop}%
\bibitem [{\citenamefont {Gervois}\ \emph {et~al.}(2002)\citenamefont {Gervois}, \citenamefont {Oger}, \citenamefont {Richard},\ and\ \citenamefont {Troadec}}]{GervoisEtAl2002}%
  \BibitemOpen
  \bibfield  {author} {\bibinfo {author} {\bibfnamefont {A.}~\bibnamefont {Gervois}}, \bibinfo {author} {\bibfnamefont {L.}~\bibnamefont {Oger}}, \bibinfo {author} {\bibfnamefont {P.}~\bibnamefont {Richard}},\ and\ \bibinfo {author} {\bibfnamefont {J.-P.}\ \bibnamefont {Troadec}},\ }\bibfield  {title} {\bibinfo {title} {Voronoi and radical tessellations of packings of spheres},\ }in\ \href@noop {} {\emph {\bibinfo {booktitle} {Computational Science --- ICCS 2002}}},\ \bibinfo {series} {Lecture Notes in Computer Science}, Vol.\ \bibinfo {volume} {2331}\ (\bibinfo  {publisher} {Springer},\ \bibinfo {year} {2002})\ pp.\ \bibinfo {pages} {95--104}\BibitemShut {NoStop}%
\bibitem [{\citenamefont {Rieser}\ \emph {et~al.}(2016)\citenamefont {Rieser}, \citenamefont {Goodrich}, \citenamefont {Liu},\ and\ \citenamefont {Durian}}]{RieserEtAl2016}%
  \BibitemOpen
  \bibfield  {author} {\bibinfo {author} {\bibfnamefont {J.~M.}\ \bibnamefont {Rieser}}, \bibinfo {author} {\bibfnamefont {C.~P.}\ \bibnamefont {Goodrich}}, \bibinfo {author} {\bibfnamefont {A.~J.}\ \bibnamefont {Liu}},\ and\ \bibinfo {author} {\bibfnamefont {D.~J.}\ \bibnamefont {Durian}},\ }\bibfield  {title} {\bibinfo {title} {Divergence of voronoi cell anisotropy vector: A threshold-free characterization of local structure in amorphous materials},\ }\href {https://doi.org/10.1103/PhysRevLett.116.088001} {\bibfield  {journal} {\bibinfo  {journal} {Phys. Rev. Lett.}\ }\textbf {\bibinfo {volume} {116}},\ \bibinfo {pages} {088001} (\bibinfo {year} {2016})}\BibitemShut {NoStop}%
\bibitem [{\citenamefont {Du}\ and\ \citenamefont {Weeks}(2024)}]{du24}%
  \BibitemOpen
  \bibfield  {author} {\bibinfo {author} {\bibfnamefont {X.}~\bibnamefont {Du}}\ and\ \bibinfo {author} {\bibfnamefont {E.~R.}\ \bibnamefont {Weeks}},\ }\bibfield  {title} {\bibinfo {title} {Rearrangements during slow compression of a jammed two-dimensional emulsion},\ }\href {https://doi.org/10.1103/PhysRevE.109.034605} {\bibfield  {journal} {\bibinfo  {journal} {Phys. Rev. E}\ }\textbf {\bibinfo {volume} {109}},\ \bibinfo {pages} {034605} (\bibinfo {year} {2024})}\BibitemShut {NoStop}%
\bibitem [{\citenamefont {Nicolas}\ and\ \citenamefont {Rottler}(2018)}]{NicolasRottler2018}%
  \BibitemOpen
  \bibfield  {author} {\bibinfo {author} {\bibfnamefont {A.}~\bibnamefont {Nicolas}}\ and\ \bibinfo {author} {\bibfnamefont {J.}~\bibnamefont {Rottler}},\ }\bibfield  {title} {\bibinfo {title} {Orientation of plastic rearrangements in two-dimensional model glasses under shear},\ }\href {https://doi.org/10.1103/PhysRevE.97.063002} {\bibfield  {journal} {\bibinfo  {journal} {Phys. Rev. E}\ }\textbf {\bibinfo {volume} {97}},\ \bibinfo {pages} {063002} (\bibinfo {year} {2018})}\BibitemShut {NoStop}%
\bibitem [{\citenamefont {Clara-Rahola}\ \emph {et~al.}(2015)\citenamefont {Clara-Rahola}, \citenamefont {Brzinski}, \citenamefont {Semwogerere}, \citenamefont {Feitosa}, \citenamefont {Crocker}, \citenamefont {Sato}, \citenamefont {Breedveld},\ and\ \citenamefont {Weeks}}]{clararahola15}%
  \BibitemOpen
  \bibfield  {author} {\bibinfo {author} {\bibfnamefont {J.}~\bibnamefont {Clara-Rahola}}, \bibinfo {author} {\bibfnamefont {T.~A.}\ \bibnamefont {Brzinski}}, \bibinfo {author} {\bibfnamefont {D.}~\bibnamefont {Semwogerere}}, \bibinfo {author} {\bibfnamefont {K.}~\bibnamefont {Feitosa}}, \bibinfo {author} {\bibfnamefont {J.~C.}\ \bibnamefont {Crocker}}, \bibinfo {author} {\bibfnamefont {J.}~\bibnamefont {Sato}}, \bibinfo {author} {\bibfnamefont {V.}~\bibnamefont {Breedveld}},\ and\ \bibinfo {author} {\bibfnamefont {E.~R.}\ \bibnamefont {Weeks}},\ }\bibfield  {title} {\bibinfo {title} {Affine and nonaffine motions in sheared polydisperse emulsions},\ }\href {https://doi.org/10.1103/physreve.91.010301} {\bibfield  {journal} {\bibinfo  {journal} {Phys. Rev. E}\ }\textbf {\bibinfo {volume} {91}},\ \bibinfo {pages} {010301(R)} (\bibinfo {year} {2015})}\BibitemShut {NoStop}%
\bibitem [{\citenamefont {Cianci}\ \emph {et~al.}(2006)\citenamefont {Cianci}, \citenamefont {Courtland},\ and\ \citenamefont {Weeks}}]{cianci06ssc}%
  \BibitemOpen
  \bibfield  {author} {\bibinfo {author} {\bibfnamefont {G.~C.}\ \bibnamefont {Cianci}}, \bibinfo {author} {\bibfnamefont {R.~E.}\ \bibnamefont {Courtland}},\ and\ \bibinfo {author} {\bibfnamefont {E.~R.}\ \bibnamefont {Weeks}},\ }\bibfield  {title} {\bibinfo {title} {Correlations of structure and dynamics in an aging colloidal glass},\ }\href {https://doi.org/10.1016/j.ssc.2006.04.039} {\bibfield  {journal} {\bibinfo  {journal} {Solid State Communications}\ }\textbf {\bibinfo {volume} {139}},\ \bibinfo {pages} {599--604} (\bibinfo {year} {2006})}\BibitemShut {NoStop}%
\end{thebibliography}
\end{document}